\documentclass{aa}
\usepackage{txfonts}
\usepackage{graphicx}
\usepackage{longtable}
\usepackage{lscape}
\newcommand{\ha}{H$\alpha$}
\newcommand{\e}{et al.\ }
\begin{document}

\title{A {\em Chandra} X-ray study of the young star cluster NGC~6231:
low-mass population and initial mass function
}

\date{Received date / Accepted date}

\author{F. Damiani\inst{1} \and G. Micela\inst{1} \and S.
Sciortino\inst{1}}
\institute{INAF - Osservatorio Astronomico di Palermo G.S.Vaiana,
Piazza del Parlamento 1, I-90134 Palermo, ITALY
}

\abstract{NGC~6231 is a massive young star cluster,
near the center of the Sco~OB1 association. While its OB
members are well studied, its low-mass population has received little attention.
We present high-spatial resolution Chandra ACIS-I X-ray data, where
we detect 1613 point X-ray sources.}
{Our main aim is to clarify global properties of NGC~6231 down to low
masses through a detailed membership assessment, and to study the
cluster stars' spatial distribution, the origin of their X-ray emission,
the cluster age and formation history, and initial mass function.}
{We use X-ray data, complemented by optical/IR data, to establish
cluster membership.  The spatial distribution of different
stellar subgroups also provides highly significant constraints on cluster
membership, as does the distribution of X-ray hardness.
We perform spectral modeling of group-stacked X-ray source spectra.}
{We find a large cluster population down to $\sim 0.3 M_{\odot}$ 
(complete to $\sim 1 M_{\odot}$),
with minimal non-member contamination, with a definite age
spread (1-8~Myrs) for the low-mass PMS stars.
We argue that low-mass cluster stars also constitute the majority of the
few hundreds unidentified X-ray sources.
We find mass segregation for
the most massive stars. The fraction of circumstellar-disk bearing
members is found to be $\sim 5$\%.
Photoevaporation of disks under the action of massive
stars is suggested by the spatial distribution of the IR-excess stars.
We also find strong \ha\ emission in 9\% of cluster PMS stars.
The dependence of X-ray properties on mass, stellar structure, and age
agrees with extrapolations based on other young clusters. The
cluster initial mass function, computed over $\sim 2$ dex in mass, has
a slope $\Gamma \sim -1.14$.
The total mass of cluster members above $1 M_{\odot}$ is $2.28 \times 10^3
M_{\odot}$, and the inferred total mass is $4.38 \times 10^3 M_{\odot}$.
We also study the peculiar, hard X-ray spectrum of the Wolf-Rayet star WR~79.}
{}


\titlerunning{The stellar population of the young cluster NGC~6231}
\authorrunning{Damiani et al.}

\maketitle

\section{Introduction}
\label{intro}

Massive stars in OB associations are easily identified as bright blue stars
standing against fainter and redder background field stars, and being
bright and observable up to large distances have been intensively
studied over decades. The low-mass stars born in the same associations,
however, have received much less attention in the past, because they are
much fainter and difficult to discern from faint field stars.
This is especially true if an OB association lies far away, low on the
Galactic plane.
The advent of X-ray
observations, however, has provided an invaluable tool to
find young low-mass stars, which are orders of magnitude brighter
in X-rays than older field stars of the same mass and spectral types.
Moreover, the sensitivity and spatial resolution of X-ray data available
today from the {\em Chandra} X-ray observatory has permitted to extend
these studies to the many regions farther than 1~kpc, previously unaccessible
(due to insufficient sensitivity and strong source confusion) to
earlier-generation X-ray observatories such as ROSAT, thus enlarging
enormously the sample of regions accessible for such detailed studies
across almost the entire mass spectrum.

We have studied in X-rays with Chandra the young cluster \object{NGC~6231},
part of
the Sco~OB1 association, containing more than 100 OB stars (and one
Wolf-Rayet star),
with the purpose of studying its population, complete down to
approximately one solar mass. These data show the existence of a very
large low-mass star population in this cluster. We are therefore able to
study in detail its spatial distribution, the cluster locus on the
color-magnitude diagram and cluster age, the percentage of stars with
T~Tauri-like properties, and the cluster initial mass function (IMF).

NGC~6231 was already observed in X-rays with XMM-Newton (Sana et
al.\ 2006a), with the detection of 610 X-ray sources; X-ray emitting
OB stars were studied by Sana et al.\ (2006b), and the lower-mass
X-ray sources by Sana et al.\ (2007). Despite these XMM-Newton data being
potentially deeper than ours (total exposure 180~ks, with larger
instrument effective area), and over a wider field, they are severely
background- and
confusion-limited, and their source catalog contains much less than one-half
sources than the new X-ray source catalog we present in this paper.

An extensive survey of the published literature on NGC~6231 was made by
Sana et al.\ (2006a), mainly optical photometric and spectroscopic
studies of cluster membership, distance, radial velocities, variability
and binarity, with main emphasis on massive stars.
Until not long ago,
optical photometric studies extending to low-mass stars
were few, with limited spatial coverage and/or relatively shallow
limiting magnitude (Sung, Bessell and Lee 1998, hereafter SBL;
Baume, V\'azquez and Feinstein 1999).
More recently, several of these limitations were overcome by the work of Sung \e
(2013a; SSB), who made deep (about $V<21$), wide-field UBVI and \ha\ photometry
in the NGC~6231 region. Moreover, the cluster has data in the VPHAS$+$ DR2
catalog (Drew \e 2014) including {\em ugri} and \ha\ magnitudes, down to a
depth comparable to SSB.
Noteworthy is the large fraction of binaries among massive stars,
as found by Raboud (1996), Garc{\'{\i}}a and Mermilliod (2001), and Sana
\e (2008).
SSB derive a distance modulus for NGC~6231 of 11.0 (1585~pc), a reddening
$E(B-V) = 0.47$, and a nearly normal reddening law with $R=3.2$.
We adopt these values for the present work.

This paper is structured as follows: Section~\ref{xobs} presents our
{\em Chandra} data on NGC~6231 and the source detection procedure; the
X-ray source cluster morphology is studied in Section~\ref{morph};
Section~\ref{ident} describes the identification of detected sources
with the available optical and near-IR catalogues;
in Section~\ref{cmd} we examine the properties of the X-ray selected
cluster population using color-magnitude and color-color diagrams;
spatial distributions of subgroups of candidate members are studied in
Section~\ref{spatial}; low-resolution ACIS X-ray spectra are studied in
Section~\ref{xspec}, and X-rays from massive OB stars are studied in
Section~\ref{himass}; eventually, the dependence of X-ray emission on
stellar properties is studied in Section~\ref{xlumfn}, and compared to
other young clusters within an evolutionary framework; a detailed cluster
IMF is computed in Section~\ref{imf}; last, we
summarize our findings in Section~\ref{concl}.
An appendix is devoted to the spatial distribution of reddening.

\section{The {\em Chandra} X-ray observations and source detection}
\label{xobs}

\begin{figure*}
\sidecaption
\includegraphics[width=12cm]{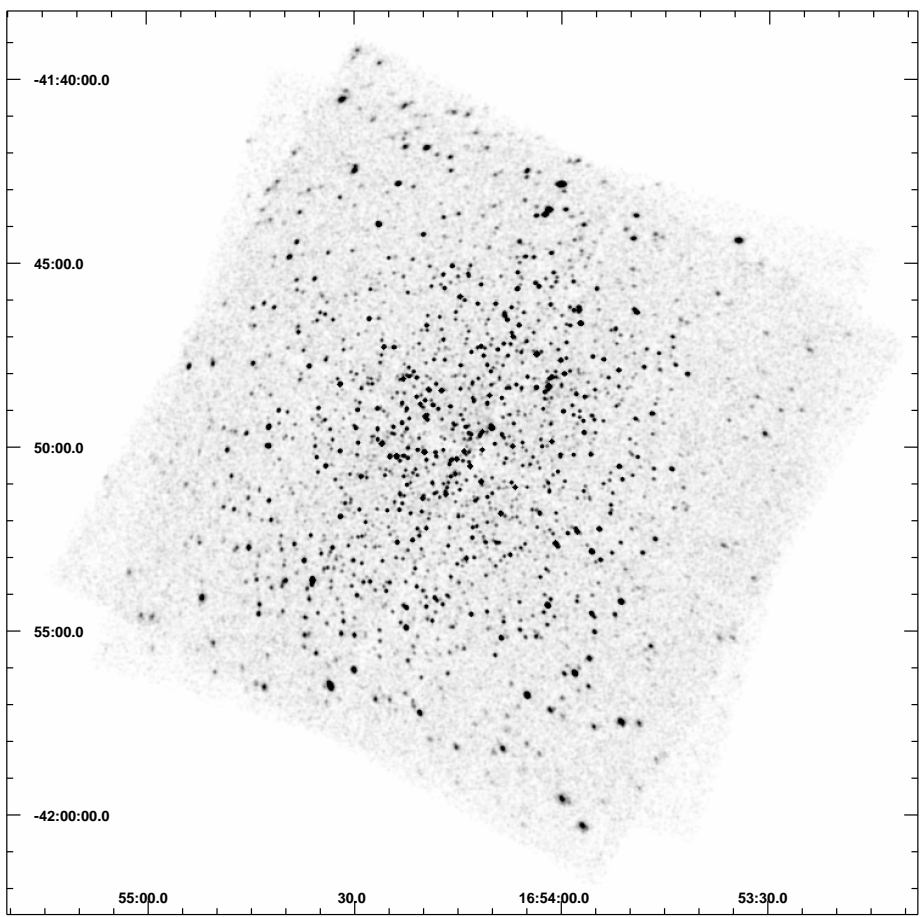}
\caption{
Combined {\em Chandra} ACIS-I image of NGC~6231, slightly smoothed
to emphasize point sources. The field of view (FOV) of each ACIS-I pointing
is $16.9^{\prime} \times 16.9^{\prime}$, corresponding to 7.81~pc at the
cluster distance.
Coordinates are equatorial (J2000).
\label{acis-smooth2}}
\end{figure*}

NGC~6231 was observed twice in X-rays with the ACIS-I detector onboard the {\em Chandra} X-ray Observatory
on 2005, July 3-4 (ObsId 5372) and 16-17 (ObsID 6291), respectively. The two pointings share the same center (aimpoint) but were performed with a
different roll angle. Effective exposure times were 76.19 and 44.39~ks, respectively, so that the total exposure time was 120.58~ks.
The combined, deep image of NGC~6231 is shown in Figure~\ref{acis-smooth2} (slightly smoothed to emphasize sources), where it is clear how crucial
is the high spatial resolution of {\em Chandra} (on-axis PSF FWHM $\sim 0.5^{\prime\prime}$) to resolve the very crowded cluster population.
At the cluster distance, the ACIS field ($16.9^{\prime}$ on a side) corresponds
to $7.81 \times 7.81$~pc.

The data were filtered to retain the energy band 0.3-8.0~keV, and the full-field
lightcurves were inspected to search for high-background periods, but none
were found. Exposure maps were computed using standard CIAO software tasks.
To these prepared datasets, we applied the source detection software
PWDetect, a wavelet-based detection algorithm developed at INAF-Osservatorio
Astronomico di Palermo\footnote{Available at
http://www.astropa.unipa.it/progetti\_ricerca/PWDetect/},
already used with success in the analysis of crowded star clusters (e.g.,
NGC~6530, Damiani et al.\ 2004, the COUP data on the Orion Nebula Cluster
(ONC), Getman et al.\ 2005, NGC~2362, Damiani et al.\ 2006a).  This algorithm
is based on its earlier ROSAT version {\em wdetect} (Damiani et al.\
1997a,b).  The PWDetect version used here is a modified one, able to
detect sources in combined datasets, thus taking fully advantage of the
deep total exposure (this version was already used to analyze the even
more varied set of {\em Chandra} data on the open cluster NGC~2516,
Damiani et al.\ 2003, and the COUP data).  The detection threshold was
chosen such as to yield ten spurious detections in the field of view
(FOV), for the given background counts. This is a more relaxed constraint
than the more usual limit of one spurious detection per field, but is
justified when the lowered threshold allows the detection of {\em more
than one hundred} additional faint sources, as it was the case here or
in the COUP Orion data.

Since the two pointings
have different roll angles, a few sources were falling in one but not in
the other (Figure~\ref{acis-smooth2}), and they were also included in the
final source list.  Overall, we detect 1613 individual X-ray sources,
listed in Table~\ref{table1}.

\begin{longtab}
\begin{landscape}
\begin{longtable}{ccccccccccccc}
\caption{\label{table1} Detected X-ray sources in NGC6231. Full table in
electronic format only.} \\
\hline  \hline
X-ray & X-ray & RA & Dec & Pos.error & Counts & Count rate & Rate error
& P$_{K-S}$ & HR1 & HR2 & $\log L_X$ & $\log L_X^q$ \\ 
no. & Id & (J2000) & (J2000) & (arcsec) & & (cts/ksec) & (cts/ksec) &
(\%) & & & (erg/s) & (erg/s) \\ 
  \hline
  1 & CXOU J165318.6-414656 & 16:53:18.62 & -41:46:56.6 & 2.80 & 44.2 & 0.651 & 0.190 &  1.8 & 0.27 & -0.29 & 30.46 &  \\ 
    2 & CXOU J165318.6-414859 & 16:53:18.62 & -41:48:59.2 & 2.95 & 32.5 & 0.610 & 0.190 &  17.7 & -0.43 & 0.17 & 30.43 &  \\ 
    3 & CXOU J165319.9-414800 & 16:53:19.91 & -41:48:00.3 & 4.48 & 34.5 & 0.539 & 0.183 &  0.0 & -0.62 & 0.61 & 30.50 & 30.26 \\ 
    4 & CXOU J165324.1-414722 & 16:53:24.16 & -41:47:22.3 & 2.21 & 68.5 & 0.797 & 0.133 &  83.6 & -0.47 & -1.00 & 30.67 &  \\ 
    5 & CXOU J165325.3-414900 & 16:53:25.33 & -41:49:00.0 & 1.87 & 49.8 & 0.544 & 0.151 &  63.0 & 0.16 & -0.46 & 30.50 &  \\ 
    6 & CXOU J165326.7-414827 & 16:53:26.80 & -41:48:27.0 & 1.87 & 47.9 & 0.534 & 0.147 &  80.2 & -0.05 & -0.32 & 30.50 &  \\ 
    7 & CXOU J165328.3-414900 & 16:53:28.33 & -41:49:00.6 & 1.72 & 50.0 & 0.548 & 0.101 &  0.0 & -0.17 & -0.00 & 30.38 & 30.16 \\ 
    8 & CXOU J165328.9-414958 & 16:53:28.98 & -41:49:58.7 & 1.23 & 13.5 & 0.142 & 0.055 &  2.7 & 0.09 & -1.00 & 29.79 &  \\ 
    9 & CXOU J165329.3-415038 & 16:53:29.40 & -41:50:38.3 & 1.18 & 18.7 & 0.196 & 0.072 &  53.4 & 0.31 & -0.50 & 29.93 &  \\ 
   10 & CXOU J165330.3-415131 & 16:53:30.33 & -41:51:31.0 & 2.46 & 22.0 & 0.229 & 0.087 &  38.2 & -0.44 & -0.11 & 30.13 &  \\ 
   11 & CXOU J165330.4-414654 & 16:53:30.45 & -41:46:54.9 & 1.77 & 54.5 & 0.595 & 0.157 &  51.6 & 0.12 & 0.04 & 30.42 &  \\ 
   12 & CXOU J165330.5-414938 & 16:53:30.59 & -41:49:38.8 & 1.08 & 98.9 & 1.049 & 0.126 &  18.2 & 0.94 & 0.40 & 30.79 &  \\ 
   13 & CXOU J165331.2-414503 & 16:53:31.21 & -41:45:03.3 & 2.02 & 36.3 & 0.500 & 0.150 &  0.1 & -0.04 & 0.00 & 30.34 &  \\ 
   14 & CXOU J165331.6-414839 & 16:53:31.65 & -41:48:39.1 & 0.98 & 34.5 & 0.368 & 0.111 &  11.1 & 0.14 & -0.90 & 30.33 &  \\ 
   15 & CXOU J165332.1-415334 & 16:53:32.19 & -41:53:34.6 & 1.77 & 47.5 & 0.530 & 0.101 &  21.1 & -0.39 & -0.37 & 30.37 &  \\ 
   16 & CXOU J165332.4-414854 & 16:53:32.50 & -41:48:54.7 & 1.97 & 40.5 & 0.430 & 0.125 &  0.0 & 0.69 & 0.45 & 30.40 &  \\ 
   17 & CXOU J165333.3-414946 & 16:53:33.40 & -41:49:46.3 & 1.92 & 42.9 & 0.435 & 0.124 &  53.8 & -0.13 & -0.49 & 30.28 &  \\ 
   18 & CXOU J165333.8-414909 & 16:53:33.87 & -41:49:09.9 & 1.33 & 14.3 & 0.150 & 0.076 &  53.9 & -0.15 & -0.83 & 29.94 &  \\ 
   19 & CXOU J165334.4-414423 & 16:53:34.42 & -41:44:23.8 & 1.08 & 305.1 & 9.404 & 0.617 &  41.4 & 0.55 & 0.19 & 31.62 &  \\ 
   20 & CXOU J165334.7-415238 & 16:53:34.72 & -41:52:38.1 & 2.12 & 32.1 & 0.332 & 0.103 &  81.1 & 0.04 & 0.04 & 30.16 &  \\ 
   \hline
\end{longtable}
\end{landscape}
\end{longtab}

\begin{longtab}
\begin{landscape}
\begin{longtable}{cccccccccccccccccc}
\caption{\label{table2} Optical/NIR identifications of X-ray sources,
and IR/H$\alpha$ excess stars. Full table in electronic format only.} \\
\hline  \hline
X-ray & SBB & $V$ & $V-I$ & H$\alpha$ & VPHAS & $r$ & $i$ &
H$\alpha$ & 2MASS & J & H & K & Grp & IR & H$\alpha$ & Mass & $L_{bol}$
\\
no. & Star & &  & SBB & Id &  & & Vphas & Id. & & & & & exc.\ &
exc.\ & $(M_{\odot})$ & $(L_{\odot})$ \\
  \hline
  1 & 12929 & 20.14 & 2.71 &  & J165318.5-414656.8 & 19.19 & 17.75 & 18.61 &  &  &  &  & 3 &  &  & 0.51 & 0.27 \\ 
    2 & 12945 & 20.38 & 3.01 &  & J165318.8-414858.6 & 19.33 & 17.83 & 18.77 &  16531817-4148569  & 15.55 &  &  & 3 &  &  & 0.39 & 0.31 \\ 
    3 & 1547 & 16.33 & 1.26 & 0.08 & J165319.8-414802.4 & 19.16 & 17.72 & 18.61 &  16531979-4148022  & 15.37 &  &  & 4 &  &  & 2.11 & 3.21 \\ 
    4 & 1616 & 13.90 & 1.14 & 0.10 & J165324.1-414722.2 & 13.44 & 12.82 & 13.20 &  16532414-4147222  & 11.99 & 11.57 & 11.46 & 3 &  &  & 2.18 & 25.67 \\ 
    5 & 13292 & 18.30 & 2.04 & -0.15 & J165325.2-414900.0 & 17.40 & 16.41 & 16.98 &  16532524-4148598  & 14.77 & 13.94 & 13.57 & 3 &  &  & 0.96 & 0.76 \\ 
    6 & 1694 & 17.31 & 1.51 & -0.03 & J165326.7-414827.2 & 16.69 & 15.93 & 16.37 &  16532673-4148271  & 14.76 & 14.16 &  & 4 &  &  & 1.88 & 1.29 \\ 
    7 & 13457 & 19.62 & 2.80 &  & J165328.3-414900.5 & 18.73 & 17.35 & 18.10 &  16532834-4148598  & 14.40 & 12.96 & 12.27 & 3 &  &  & 0.47 & 0.49 \\ 
    8 &  &  &  &  & J165329.0-414958.6 & 20.10 & 18.30 & 19.43 &  16532901-4149596  & 14.96 & 14.07 & 13.75 & 3 &  &  &  &  \\ 
    9 &  &  &  &  & J165329.5-415037.7 & 19.78 & 18.22 & 18.99 &  16532933-4150372  &  &  & 13.51 & 3 &  &  &  &  \\ 
   10 & 13561 & 19.20 & 2.40 & 0.09 & J165330.4-415130.2 & 18.25 & 17.06 & 17.63 &  16533030-4151307  & 14.93 & 13.63 & 13.12 & 3 &  &  & 0.65 & 0.48 \\ 
   11 & 13566 & 19.32 & 2.49 &  & J165330.5-414654.7 & 18.58 &  & 18.01 &  16533046-4146543  & 14.87 & 13.89 & 13.54 & 3 &  &  & 0.60 & 0.48 \\ 
   12 & 1805 & 17.78 & 1.45 & 0.00 & J165330.5-414939.4 & 17.19 & 16.42 & 16.85 &  16533046-4149394  & 15.27 &  &  & 4 &  &  & 1.98 & 0.80 \\ 
   13 & 13610 & 19.72 & 2.66 &  & J165331.2-414503.1 & 18.86 & 17.51 & 18.29 &  16533125-4145033  & 15.25 & 14.32 & 13.94 & 3 &  &  & 0.53 & 0.39 \\ 
   14 & 1841 & 16.87 & 1.74 & -0.05 & J165331.6-414839.2 & 16.16 & 15.28 & 15.78 &  16533163-4148391  & 13.91 & 13.19 & 12.99 & 3 &  &  & 1.33 & 2.07 \\ 
   15 & 13665 & 18.90 & 2.38 & -0.11 & J165332.2-415334.5 & 17.78 & 16.60 & 17.29 &  16533214-4153343  & 14.79 & 13.88 & 13.76 & 3 &  &  & 0.66 & 0.62 \\ 
   16 & 13672 & 19.88 & 1.95 &  & J165332.9-414856.9 & 20.16 & 18.95 & 19.51 &  16533237-4148548  &  & 14.89 & 14.41 & 4 &  &  & 1.05 & 0.16 \\ 
   17 & 13738 & 19.43 & 2.58 &  & J165333.4-414946.4 & 18.49 & 17.22 & 17.92 &  16533340-4149463  & 15.18 & 14.34 & 14.08 & 3 &  &  & 0.56 & 0.47 \\ 
   18 &  &  &  &  & J165333.8-414909.3 & 19.52 & 17.88 & 18.78 &  16533379-4149092  & 15.95 &  &  & 3 &  &  &  &  \\ 
   19 &  &  &  &  & J165334.5-414419.8 & 20.48 & 19.07 &  &  16533407-4144224  & 15.80 & 14.75 & 14.32 & 4 &  &  &  &  \\ 
   20 & 13822 & 19.28 & 2.34 & -0.08 & J165334.8-415238.0 & 18.39 &  & 17.86 &  16533481-4152381  & 15.33 &  &  & 3 &  &  & 0.68 & 0.42 \\ 
   \hline
\end{longtable}
\end{landscape}
\end{longtab}

\section{Cluster size and morphology}
\label{morph}

\begin{figure}
\resizebox{\hsize}{!}{
\includegraphics[bb=5 10 485 475]{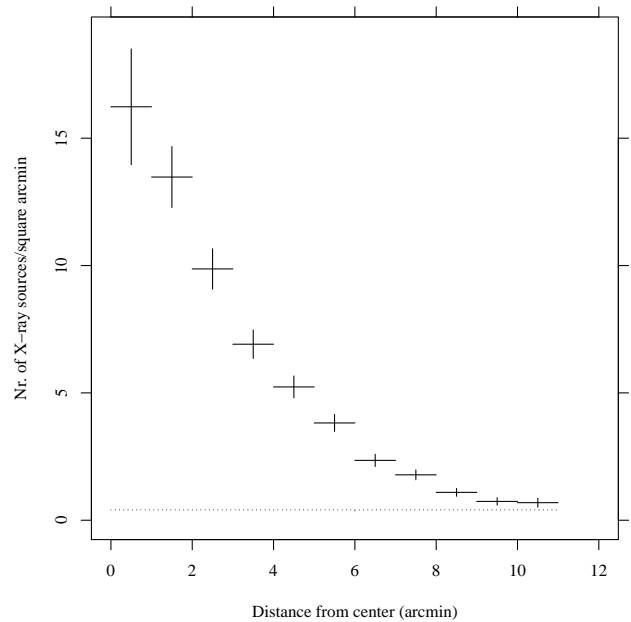}}
\caption{Surface density of X-ray sources detected in the ACIS FOV, with
more than 20~X-ray counts. The dotted horizontal line is the average
surface density of a reference galactic-plane field with nearly the same
exposure time as the NGC~6231 field.
\label{clust-profile3}}
\end{figure}

\begin{figure}
\resizebox{\hsize}{!}{
\includegraphics[bb=5 10 485 475]{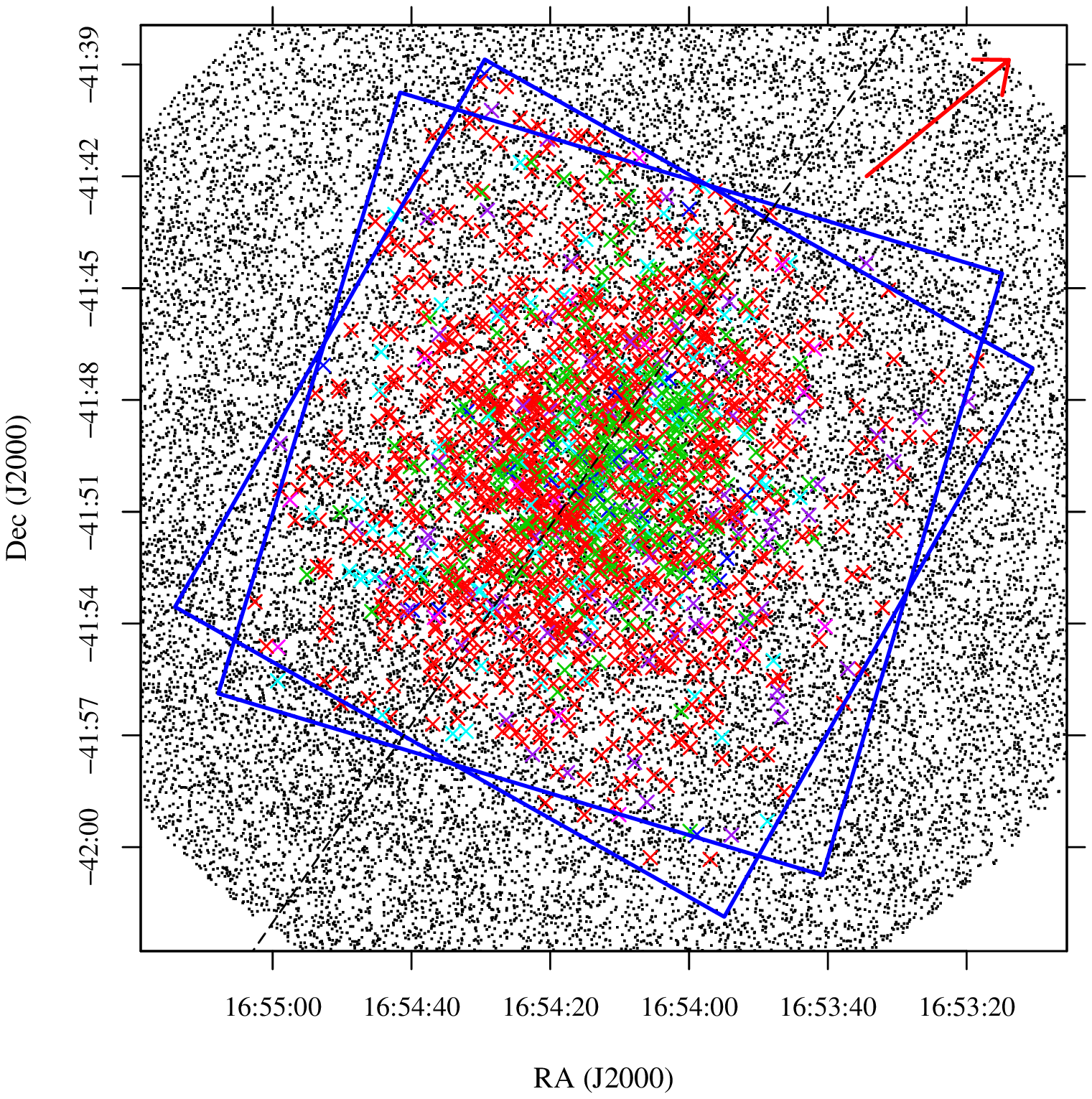}}
\caption{Spatial distribution of all detected X-ray sources (crosses),
and of stars from the SSB optical catalogue
(dots) within 15$^{\prime}$ of cluster center.
The two tilted blue squares are the two ACIS-I FOVs.
The dashed line running
almost diagonally throughout the figure is the cluster symmetry axis
obtained from our principal-component analysis. The Galactic Plane
direction is here approximately towards South-East.
The big red arrow at top right indicates galactic North.
Color codes for X-ray detected sources (crosses) are: blue for OB stars,
cyan for A-F (including some B-type) stars,
red for low-mass pre-main-sequence stars,
purple for field MS stars, magenta for foreground stars, and green
for optically unidentified sources.
\label{spatial-map-xopt}}
\end{figure}

The very large number of X-ray sources in the {\em Chandra} image,
compared to the average galactic-plane field, and their clustered
distribution, strongly suggest that most X-ray sources are indeed
cluster members.  Very young stars are known to emit copious X-rays,
with X-ray luminosities typically $10^3-10^4$ times higher than (older)
field stars. X-ray detection is therefore an important, if not conclusive,
membership criterion, which we will use in combination with the optical
and infrared data for these stars to find a list of good candidate
members of the cluster.

The cluster has (almost) a circular symmetry, and we have studied its
radial profile by counting detected sources in annuli, centered on the
median position of all X-ray sources: ($\alpha,\delta$) $=$ (16:54:14.4,
--41:49:54.8).  We consider only detections with
more than 20 counts, i.e.\ detectable across the whole FOV, to be free
from off-axis dependent sensitivity effects (mainly due to the strongly
varying PSF).  The cluster profile is shown in Figure~\ref{clust-profile3}:
the source density falls down by a factor of 2 within 3~arcmin (1.39~pc)
from center, and becomes compatible with an average Galactic-plane value
beyond 9-10~arcmin. This characteristic radius is 48\% larger than the
corresponding radius of NGC~2362, determined with exactly the same method
and above nearly the same count-rate threshold (Damiani et al.\ 2006a).

To estimate an appropriate Galactic-plane X-ray source density above
the same threshold, in the direction of NGC~6231, we have selected from
the {\em Chandra} archive a reference pointing on the Galactic plane,
purposely free from extended sources or star clusters, and as close as
possible in angular distance from galactic-center, and exposure time,
to our pointings.
One such dataset is ObsId~2298 (PI: K.Ebisawa), with exposure 98.7~ks,
at galactic coordinates $(l,b) = (28.56,-0.02)$ (NGC~6231 lies at
$(343.46,1.17)$). The source density above 18 counts (i.e.\ 20 counts
scaled to the exposure-time ratio) is shown as a dotted horizontal
line in Figure~\ref{clust-profile3}. This galactic plane density value
is only approximately appropriate, since it depends strongly on the
actual pointing direction, and decreases with distance from the Galactic
Center and Galactic Plane (which are not the same for NGC~6231 and the
reference field).  A corresponding reference field for NGC~2362 showed
a distinctly lower source density than ObsId~2298 (see Damiani et al.\
2006a).

Upon close inspection, the distribution of X-ray sources in NGC~6231
is not exactly circular, but slightly elongated. We applied
a principal-component analysis to source positions (limited to
off-axis less than 7~arcmin, to be unaffected by the square ACIS-I
shape), and found a symmetry axis inclined by 35~degrees from North
(Figure~\ref{spatial-map-xopt}), namely approximately normal to the
Galactic Plane.  The rms distances of sources from cluster center along and
across such axis differ by 15\%. There is however no apparent dependence of
stellar properties on this elongated morphology: ages and masses (to be
studied below) are
distributed identically on opposite sides along the symmetry axis.
This elongation might be related to the origin of the cluster itself:
Rees and Cudworth (2003) suggested that the passage through the galactic
disk of the globular cluster NGC~6397 might have triggered formation of
NGC~6231.

\section{Optical and IR identifications of X-ray sources}
\label{ident}

We have identified our X-ray detections with stars in the SSB catalog,
in the VPHAS$+$ catalog, and the 2MASS catalog. Each of these catalogues
covers fully our combined X-ray FOV.
Source identification was first made with 2MASS (as astrometric
reference), using individual position errors computed by PWDetect for
X-ray detections and catalogued values for 2MASS sources\footnote{
Following the 2MASS Explanatory Supplement (Cutri et al.\ 2003), we
considered only 2MASS sources unaffected by
strong photometric uncertainties, and discarded those with flags
{\it ph\_qual=`E'}, {\it `F'}, {\it `U'} or {\it `X'}, or with
{\it cc\_flg=`p'}, {\it `d'} or {\it `s'}.}.
Matches were accepted within a distance of 4~$\sigma$ times the combined
position uncertainty. 
In this way we found and corrected a systematic
shift between X-ray- and 2MASS positions,
of $0.08^{\prime\prime}$ in R.A., and $0.12^{\prime\prime}$ in Declination.
After accounting for these corrections, we find 1087 2MASS
identifications for our X-ray sources.
Among them, the estimated number of spurious identifications is 126.

We proceeded in an analogous way in matching X-ray sources with the
VPHAS$+$ and SSB catalogues. VPHAS$+$ identifications are 1164, while
SSB yielded 1081 identifications\footnote{We made some sub-arcsecond
adjustments to the SSB positions, using second-degree polynomials in RA
and Dec, since the SSB positions exhibited non-random,
position dependent offsets with respect to X-ray positions; the same
effect was evidently absent using VPHAS$+$ and 2MASS positions.}.
The estimated numbers of spurious X-ray identifications are 90 for VPHAS$+$
stars and 83 for SSB stars.
The three catalogues, although spatially complete across our X-ray FOV,
have each its own incompleteness, so that the overlap of the three
identification lists is not complete (nor is any of the lists simply a
subset of any other list). There are 161 SSB X-ray identifications
without a VPHAS$+$ counterpart, and 151 without a 2MASS counterpart.
This is due mostly to incompleteness of the latter two catalogues at the
bright end: NGC~6231 is a rich cluster, with its brightest stars having
$V \sim 5$, or 15-16 magnitudes above the faint limit of those
catalogues. On the other hand, we have checked that the SSB catalogue
depth is not uniform across the FOV: the southwest quadrant is $\sim
1.5$~magnitudes shallower than the rest of the region.
As a result, 249 X-ray sources with a VPHAS$+$ identification have no
counterpart in the SSB catalogue, and 226 have also no counterpart in
2MASS. Finally, among the 2MASS identifications of X-ray sources, 179
have no VPHAS$+$ counterpart, and another 179 have no SSB counterpart.
Overall, the number of X-ray sources with at least a counterpart in any
of the optical/2MASS catalogues is 1374; only 239 X-ray sources remained
without any identification.
Results from this identification procedure are given in Table~\ref{table2}.

\section{Color-magnitude and color-color diagrams}
\label{cmd}

\subsection{Stellar groups}
\label{groups}

\begin{figure}
\resizebox{\hsize}{!}{
\includegraphics[bb=5 10 485 475]{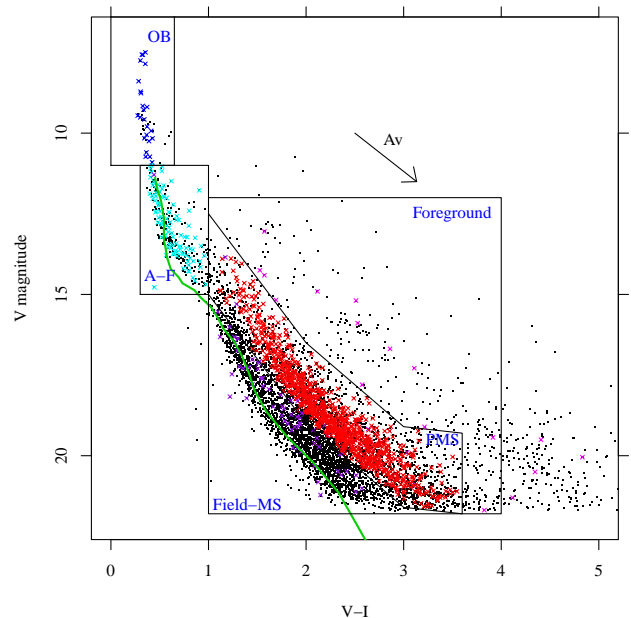}}
\caption{The $(V,V-I)$ color-magnitude diagram of all
optical stars in the NGC~6231 ACIS FOV.  Crosses are X-ray sources,
with color codes as in Fig.~\ref{spatial-map-xopt},
while dots are X-ray undetected stars.
The solid green line is the ZAMS at the cluster distance and reddening.
The black polygons are used to define stellar groups, as labeled.
The cluster extinction vector is also shown.
\label{v-vi-xray}}
\end{figure}

In Figure~\ref{v-vi-xray} we present a $(V,V-I)$ color-magnitude diagram
(CMD), of both X-ray detected and undetected stars in the ACIS FOV,
using the SSB photometry.
The solid line is the Siess \e (2000) zero-age main sequence (ZAMS) using
the Kenyon and Hartmann (1995; KH95) color transformation.
This reproduces approximately the lower envelope of field-star datapoints. Very
conspicuous is the lack of faint blue stars below it, which indicates a
rapid rise of extinction with distance. In the case of an extinction ``wall'',
as is the case behind the cluster NGC~6530 (Prisinzano \e 2005), the ZAMS
shape describes accurately the lower envelope of field-star datapoints in the
CMD. The characteristics ZAMS shape becomes instead unrecognizable when
extinction increases more slowly with distance.
In the case of NGC~6231 the agreement between the ZAMS shape and observed
datapoints is limited, so we cannot unambiguously determine the presence of
an extinction wall just behind the cluster.
Nevertheless, evidence for a sharp extinction rise somewhere behind
the cluster and its neighborhood, based on IR color distributions,
is discussed in Appendix~\ref{extinct-map}.

\begin{figure}
\resizebox{\hsize}{!}{
\includegraphics[bb=5 10 485 475]{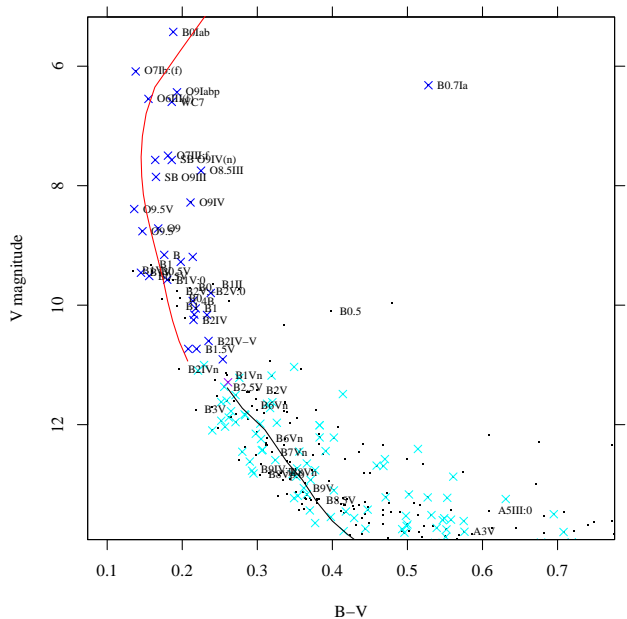}}
\caption{The $(V,B-V)$ color-magnitude diagram of bright stars in
NGC~6231. Symbols are as in Figure~\ref{v-vi-xray}.
When available from SIMBAD, spectral types are indicated.
The black line is the KH95 ZAMS, while the red line is a 3~Myr
isochrone from Ekstr\"om \e (2012).
\label{v-bv-xray-himass}}
\end{figure}

From the CMD in Figure~\ref{v-vi-xray}, it is clear that the majority of X-ray
sources consist of many tens of bright stars on the ZAMS,
and hundreds stars in a 2-3 mag wide band above the low-mass main sequence
(MS) at $V-I>1$, which is natural to identify as
the cluster pre-main-sequence (PMS) band, as also found by Sana et al.\ (2007).
Accordingly, we define five main star groups from their position in the CMD
of Fig.~\ref{v-vi-xray}, as indicated: the OB (main-sequence or evolved)
stars, the A-F stars which have just arrived on the main sequence (or
are about to arrive), the PMS stars making the bulk of the X-ray
detections, and Field-MS/Foreground stars as X-ray detections distinctly
below/above the PMS band.
At the bright end, several stars in the SSB catalogue lack a $V-I$
color value but have a $B-V$ value; we included these latter in the OB
group if $B-V<0.65$ and $V<11$.
The upper main sequence of NGC~6231 is better shown in the $(V,B-V)$
CMD of Figure~\ref{v-bv-xray-himass}, where only
stars brighter than $V=14$) are included, with their spectral types
indicated (when available from SIMBAD).
All stars brighter than $V \sim
9.3$ are detected in X-rays, including 4 OB supergiants and the
Wolf-Rayet star \object{HD~152270} (WR~79, X-ray \#1008).
Spectral types of stars fainter than $V \sim 12$ are increasingly rare,
so there is little detailed information available
for individual stars in NGC~6231.

\begin{figure}
\resizebox{\hsize}{!}{
\includegraphics[bb=5 10 485 475]{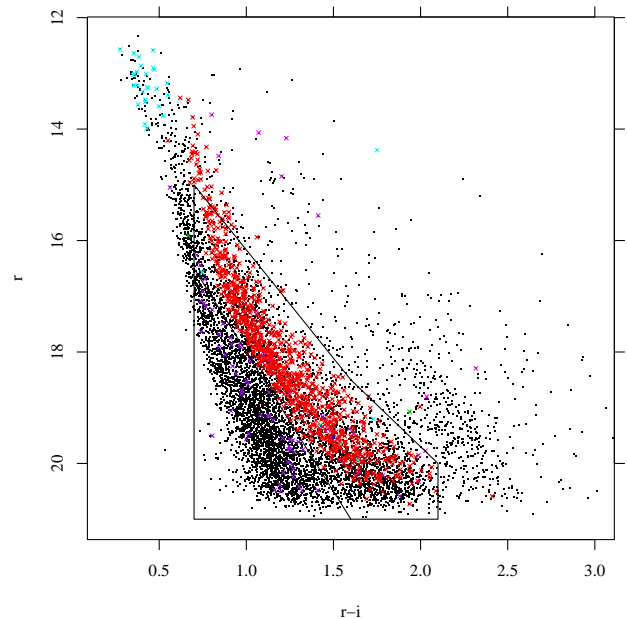}}
\caption{VPHAS$+$ $(r,r-i)$ CMD. Symbols as in
Fig.~\ref{spatial-map-xopt}. The adopted definition of additional
VPHAS$+$ stars in the PMS and Field-MS groups is indicated by the polygons.
\label{vphas-r-ri}}
\end{figure}

At the faint end of the complete CMD, the grouping based on
the SSB catalogue was complemented with the VPHAS$+$ catalogue, using
the $(r,r-i)$ CMD shown in Figure~\ref{vphas-r-ri}.
A sixth group is also defined, which comprises all X-ray
sources not identified with an optical star.
Overall we classify 31 X-ray sources in the OB group, 95 in the A-F
group, 1035 in the PMS group, 80 in the Field-MS group, 18 in the
foreground, and 354 as optically unidentified. The nature of these latter X-ray
sources, and the completeness of our X-ray catalogue at the lowest masses
is discussed below.

\begin{figure}
\resizebox{\hsize}{!}{
\includegraphics[bb=5 10 485 475]{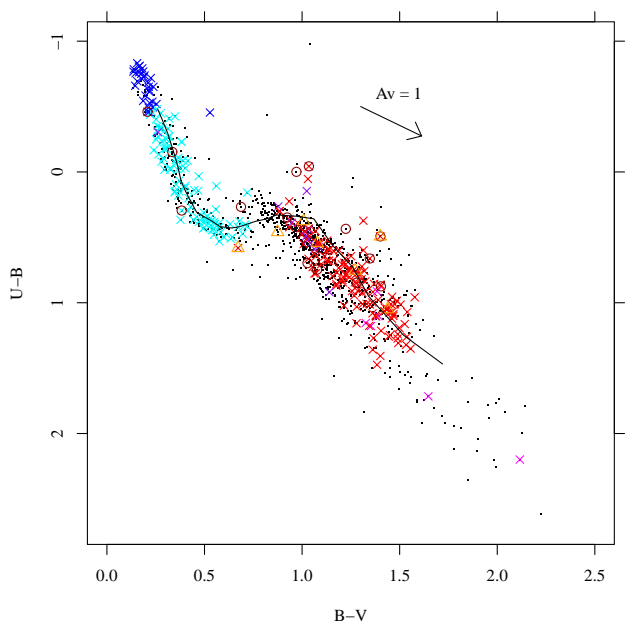}}
\caption{$(U-B,B-V)$ color-color diagram for NGC~6231. Symbols are as in
Figure~\ref{v-vi-xray}.
Orange triangles indicate IR-excess stars, and brown circles \ha-excess
stars, see below.
\label{ub-bv-xray}}
\end{figure}

We use the $(U-B,B-V)$ optical color-color diagram of
Figure~\ref{ub-bv-xray} to estimate whether there is a different
extinction affecting massive and low-mass stars, and whether stars with
strong UV excesses are present in the cluster (as is typical of e.g.\
strongly accreting T~Tauri stars). It turns out that the same average reddening
value describes well nearly all X-ray detected stars, with most of the
exceptions having large photometric errors. While a definite spread,
exceeding photometric errors, exists for massive stars (in agreement with
SSB), the spread for
low-mass stars can be accounted for by the photometric errors. In any
case, the fact that the average reddening is the same for all stars
implies that its nature is interstellar (foreground), not circumstellar.
Moreover, there is no conclusive evidence of UV excesses in a significant
number of cluster stars, including those showing \ha\ emission or IR
excesses, to be defined below, which are diagnostics of accretion and
circumstellar disks in PMS cluster members.

\subsection{Cluster age and age spread}
\label{ages}

\begin{figure}
\resizebox{\hsize}{!}{
\includegraphics[bb=5 10 485 475]{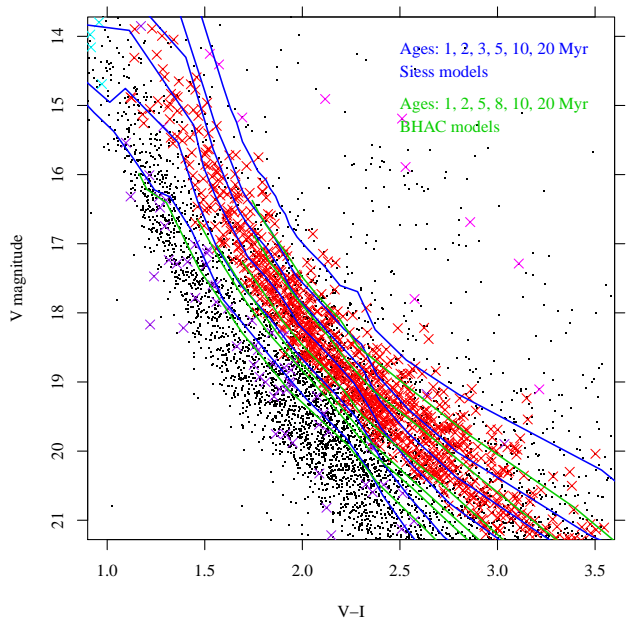}}
\caption{PMS band of the cluster CMD with superimposed BHAC and Siess \e (2000)
isochrones of different ages, as labeled.
Symbols as in Fig.~\ref{v-vi-xray}.
\label{v-vi-xray-siess-bhac}}
\end{figure}

We have compared the positions of the PMS cluster members in the CMD
with available model isochrones. The set of models by Baraffe \e (2015;
BHAC), shown in Fig.~\ref{v-vi-xray-siess-bhac} with green lines,
describe well the observed cluster locus, with inferred ages and spread
independent of mass. The BHAC models are limited to stars with $M<1.4
M_{\odot}$. Therefore, we have also considered the older Siess \e (2000)
isochrones, also shown in Fig.~\ref{v-vi-xray-siess-bhac} with blue lines,
which perform less well in the same mass range (with color-depending
inferred ages for our stars). We have therefore
adopted preferentially the BHAC isochrones as reference, and the Siess
\e ones only for $M>1.4 M_{\odot}$. With these choices, the bulk of
cluster PMS stars fall in the age range
1-8~Myr, in good agreement with the ages found by SSB (4-7~Myr for the
massive stars, 1–7~Myr for low-mass PMS stars).

\begin{figure}
\resizebox{\hsize}{!}{
\includegraphics[bb=5 10 485 475]{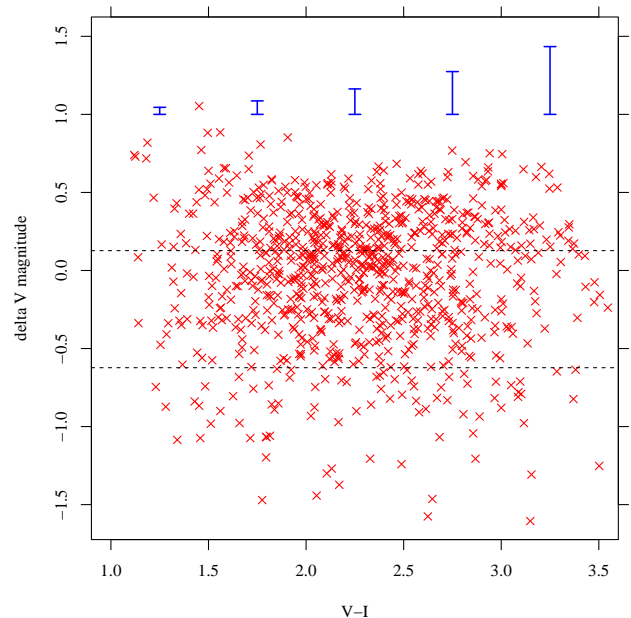}}
\caption{$V$ magnitude residuals $\Delta V$ vs.\ $V-I$ color, for X-ray detected
PMS cluster stars (red crosses). Also shown are the average total errors on
$\Delta V$, for comparison (blue segments).
\label{delta-v-age}}
\end{figure}

We have studied if the spread in the PMS band of NGC~6231 is significant
with respect to errors, and the age spread this might imply. To better
quantify such a spread, $V$ magnitude differences were computed with
respect to the BHAC 5-Myr isochrone, but this yielded color-dependent
residuals when examined in detail. Therefore, we proceeded by using an
empirical running-mean fit to the PMS cluster stars, which yielded
color-independent residuals $\Delta V$, as shown in
Figure~\ref{delta-v-age} (red crosses). In the same Figure we show with
blue segments the (color-dependent) averages of total error
on $\Delta V$, computed as
$d V = \sqrt{(\delta V)^2 + \left(\left|\frac{\partial V}{\partial (V-I)}\right|
\delta (V-I)\right)^2}$, where $\delta V$ and $\delta (V-I)$ are the
photometric errors given by SSB. This formulation is necessary since
even if (hypothetically) the $V$ magnitude error $\delta V$ were equal
to zero, the $V-I$ error would still produce a broadened sequence, and
therefore a spread in $V$, to an amount proportional to the local slope
of the cluster sequence, here being $\frac{\partial V}{\partial (V-I)} \sim
3.37$.
Figure~\ref{delta-v-age} shows that, except for the reddest cluster
stars, total errors on $\Delta V$ are much smaller than the observed
spread, which is therefore real, and attributable to an age spread but
also to some extent to binaries.

\begin{figure}
\resizebox{\hsize}{!}{
\includegraphics[bb=5 10 485 475]{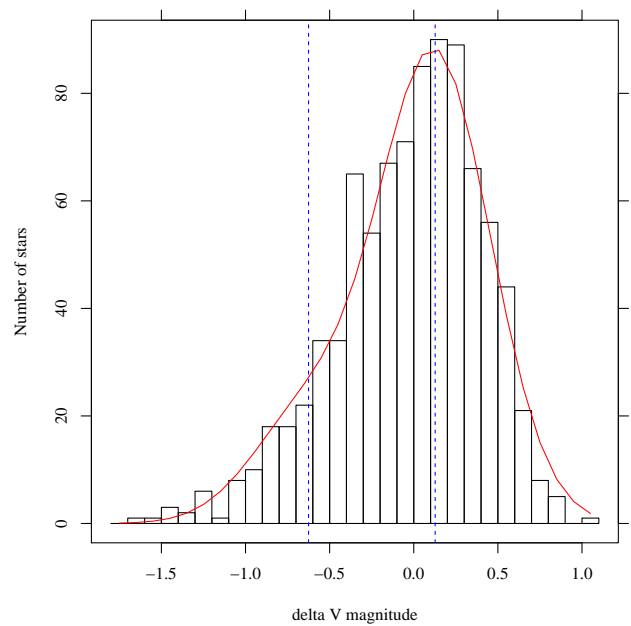}}
\caption{Distribution of $\Delta V$ for cluster PMS stars, with best-fit
model (red). Dashed blue lines indicate the centers of the best-fit
Gaussians.
\label{binary-seq2}}
\end{figure}

A histogram showing the distribution of $\Delta V$ is shown in
Figure~\ref{binary-seq2}. It is clear that the distribution is
asymmetrical, with a more extended tail at negative values. This might
be due to the time-varying star-formation rate in the cluster, in the
absence of binaries. However, postulating the absence of binaries is
highly unlikely (especially since a large binary fraction of at least
68\% was found for massive stars by Sana \e 2008). A low-mass binary sequence is
found in the 5~Myr old cluster NGC2362 by Moitinho \e (2001), thanks to its
exceptionally clear CMD. We have therefore
modeled the $\Delta V$ distribution using two Gaussians (for single and
binary stars respectively), with the same width $\sigma$, separated by a fixed 
interval of 0.752 magnitudes, and independent normalizations. The
best-fit model, shown with the red line in the Figure, reproduces well
the observed histogram: its parameters are $\sigma=0.33$~mag, and
normalization ratio 0.235 (binaries/singles). Such binary fraction is
much lower than that of massive stars, but it must be remarked that it
refers to (near) equal-mass binaries only, so it is actually a lower
limit. The 0.33~mag net spread on $\Delta V$ (still much larger than
errors on $\Delta V$, as derived above) is instead attributed to intrinsic
age spread among the cluster PMS stars. Compared to the BHAC isochrones,
this amounts to approximately a factor of two on either direction
($1\sigma$) from median, so that the bulk of PMS stars is derived to
have ages in the range 1.5-7~Myr.

\subsection{\ha\ emission stars}
\label{halpha}

\begin{figure}
\resizebox{\hsize}{!}{
\includegraphics[bb=5 10 485 475]{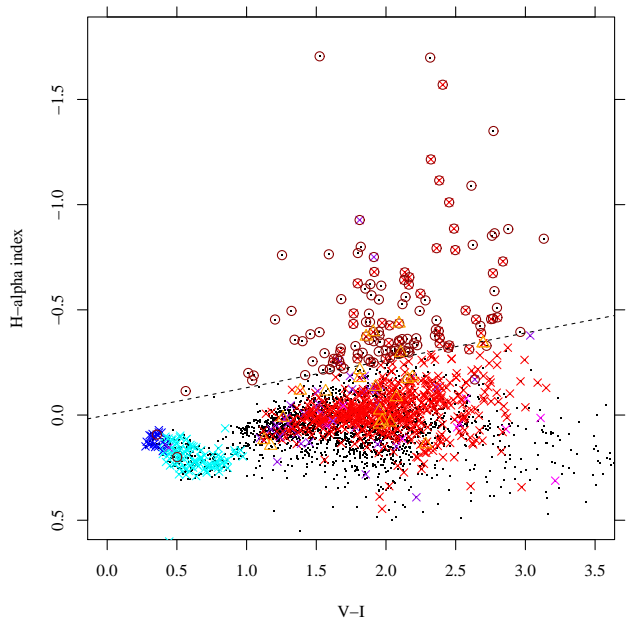}}
\caption{\ha\ index from SSB vs.\ $V-I$ color. 
Orange triangles indicate IR-excess stars (see below), and brown circles
\ha-excess stars; other symbols as in Fig.~\ref{v-vi-xray}.
\label{sung-halpha}}
\end{figure}

SBL tried to find low-mass NGC~6231 stars on the basis of their enhanced
\ha\ emission, typical of T~Tauri stars, but found only a relatively
small number of stars, with an inferred low-mass depleted IMF. This was later
rectified by the work of SSB, who found a much richer low-mass
population. The \ha\ photometry presented by SBL, and to a deeper
limiting magnitude by SSB, keeps its value as a tool to study the
fraction of PMS cluster stars showing classical T~Tauri star (CTTS)
characteristics, as opposed to PMS stars exhibiting only weak emission
lines (weak-line T~Tauri stars - WTTS). Moreover, some CTTS with thick
circumstellar envelopes might remain undetected in X-rays, so that a
strong \ha\ emission complements X-rays as a cluster membership
indicator. The selection of CTTS using the SSB photometric data is
conveniently done (as these authors already did) using a diagram of
their \ha\ index vs.\ $V-I$ color, which we reproduce in
Figure~\ref{sung-halpha} with the addition of our X-ray detections.
The dashed line is a nominal limit to separate CTTS (all stars above it)
from WTTS and other non-emission stars. We have set this limit in a more
conservative way than SSB. As just mentioned, not all CTTS selected
using this diagram are X-ray sources (crosses).

\begin{figure}
\resizebox{\hsize}{!}{
\includegraphics[bb=5 10 485 475]{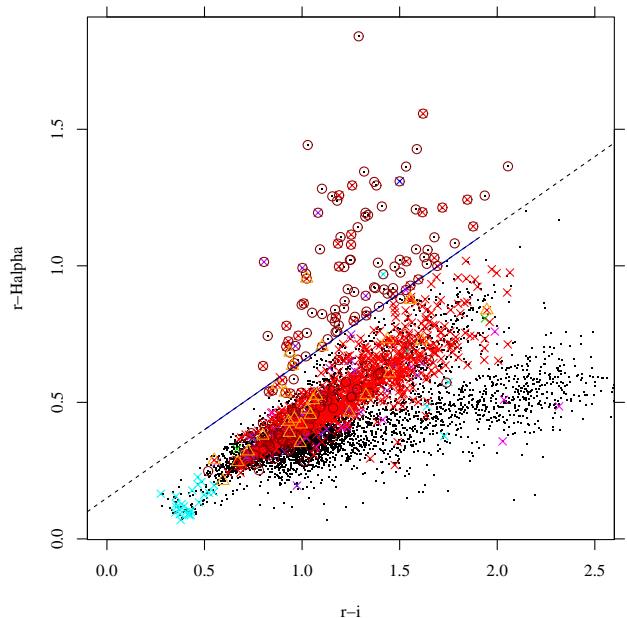}}
\caption{VPHAS$+$ index $r-H\alpha$ vs.\ color $r-i$.
Symbols as in Fig.~\ref{sung-halpha}.
\label{vphas-excess}}
\end{figure}

A useful complement to the SSB photometry is the VPHAS$+$ photometry,
also including an \ha\ emission index $r-H\alpha$. A diagram showing
this index vs.\ color $r-i$ (also from VPHAS$+$) is in
Figure~\ref{vphas-excess}. The nominal separation between CTTS and
non-emission stars shown in the Figure is taken from Kalari \e (2015).
Therefore, we have defined CTTS in NGC~6231 as the union of the CTTS
samples derived from SSB and VPHAS$+$ catalogues (203 stars in the ACIS
FOV, of which 98 in the PMS locus). The total number of PMS stars
(either X-ray or \ha-selected) becomes 1075, and the CTTS/total number
ratio is 9.1\%. Most of the \ha-selected stars falling outside the cluster PMS
band lie in the Field-MS band (101 stars), which is a somewhat surprisingly
large number of stars for a sample of foreground main-sequence field
stars. There is still the possibility that some (maybe most) of these
near-MS \ha-excess stars are true PMS NGC~6231 members, with peculiar
optical colors making them appear bluer, or at lower luminosity, than
the bulk of cluster PMS stars. An analogous situation was described by
Damiani \e (2006b) in NGC6530 and Guarcello \e (2010a) in NGC6611,
and cannot be discarded here.
If the near-MS \ha-emission stars in NGC~6231 are PMS members,
their X-ray detection rate (7/101 stars) is remarkably low,
compared to the X-ray detection rate of \ha-emission stars in the PMS
band (58/98 stars).
Also in NGC~2264 several \ha-emitting candidate members are found near
or even below the MS, and typically lack X-ray emission, perhaps because
of absorption by a nearly edge-on disk (Sung \e 2004).
However,
since we lack for the moment a definite indication
that the near-MS \ha-emission stars are indeed members, we
conservatively exclude them from further considerations in the
following.

\begin{figure}
\resizebox{\hsize}{!}{
\includegraphics[bb=5 10 485 475]{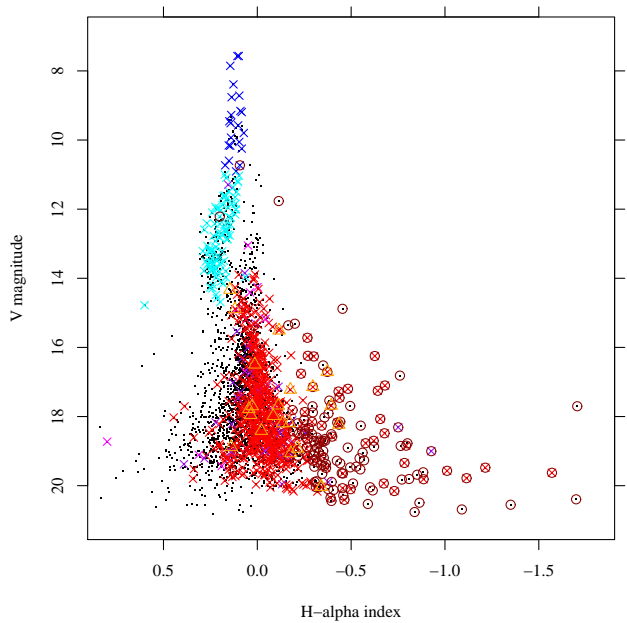}}
\caption{Index \ha\ from SSB vs.\ $V$ magnitude.
Symbols as in Fig.~\ref{sung-halpha}.
\label{v-rha-xray}}
\end{figure}

Also worth consideration is a diagram of index \ha\ (from SSB) vs.\ $V$
magnitude (Figure~\ref{v-rha-xray}). Besides the \ha-excess stars
already selected, this diagram shows
a discontinuity in the distribution of X-ray sources
around $V=14$, which is not found in
other clusters (see e.g.\ the analogous diagram for NGC6530, Figure~14
in Damiani et al.\ 2004). By
comparison with the $(V,V-I)$ diagram, we see that near $V=14$ we find the bulk
of A stars just settled on the main sequence, after crossing in a
relatively short time the diagram horizontally. Because of this, in
Figure~\ref{v-rha-xray} near
$V=14$ a color gap separates a rich group of A stars (with deep
H$\alpha$ absorption lines, and lowest values of \ha\ index), from G
stars with weak (photospheric) H$\alpha$ lines. At younger ages (e.g.\
2-3~Myr as NGC~6530) A-type stars have not yet arrived on the
main-sequence, but are found, in a smaller number, on their PMS
radiative tracks, during a rapid transition to
become main-sequence B stars. Therefore, a diagram like that of
Figure~\ref{v-rha-xray} can be another useful age indicator for a
cluster, where the \ha\ index discontinuity increases with age from 2 to
10~Myrs. This allows us to trust more confidently the NGC~6231 isochronal
age derived above.

\subsection{IR diagrams}
\label{nir}

\begin{figure}
\resizebox{\hsize}{!}{
\includegraphics[bb=5 10 485 475]{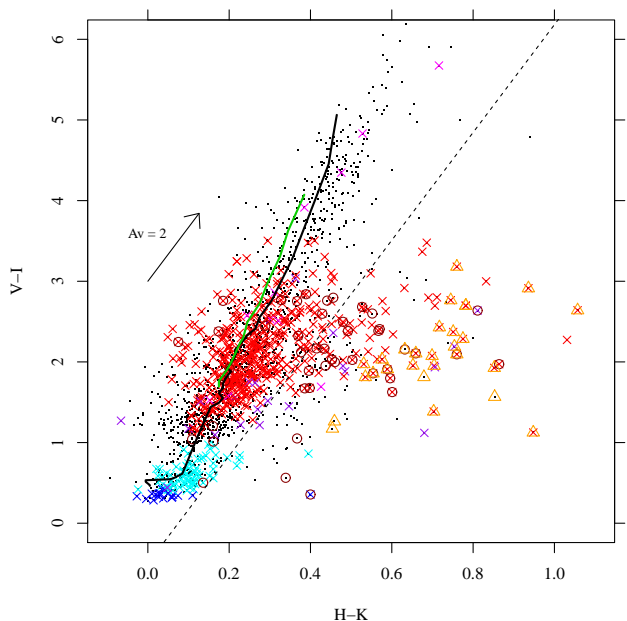}}
\caption{Mixed optical-IR $(V-I,H-K)$ color-color diagram. Symbols are as in
Figure~\ref{v-vi-xray}.
The blue (green) solid line is a 5~Myr Siess (BHAC) isochrone for mass
$M>0.1 M_{\odot}$, the dashed line is our
adopted limit to select IR-excess stars.
Symbols as in Fig.~\ref{sung-halpha}.
\label{vi-hk}}
\end{figure}

\begin{figure}
\resizebox{\hsize}{!}{
\includegraphics[bb=5 10 485 475]{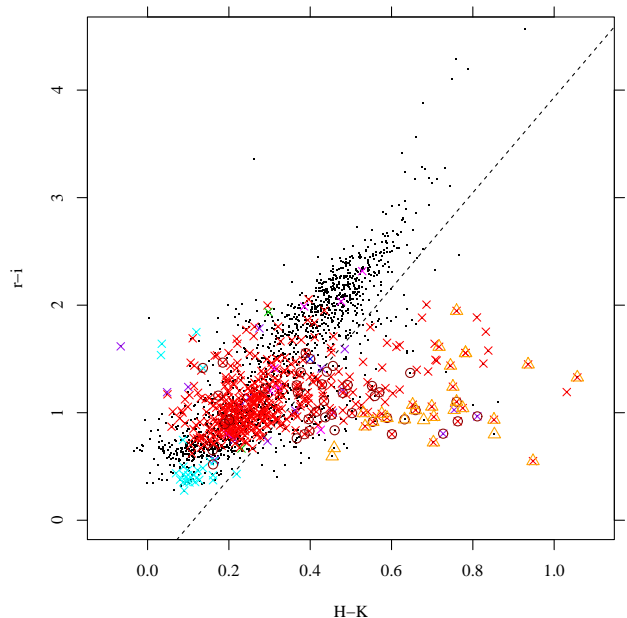}}
\caption{Optical-IR $(r-i,H-K)$ color-color diagram using VPHAS$+$
colors. Symbols are as in Figure~\ref{v-vi-xray}. The dashed line is our
adopted limit to select IR-excess stars (orange triangles).
Symbols as in Fig.~\ref{sung-halpha}.
\label{ri-hk}}
\end{figure}

The 2MASS photometry provides us with additional useful
information on the studied stars.
We first consider
mixed optical-IR color-color diagrams, such as the $(V-I,H-K)$
diagram of Figure~\ref{vi-hk}. The usefulness of this type of diagrams
for selecting stars with near-IR excesses
was previously demonstrated in our study of the massive young cluster
NGC~6530 (Damiani et al.\ 2006b): many IR-excess stars can be found from
these mixed diagrams, which are not selected as such from the usual $(J-H,H-K)$
diagram, because their IR colors alone mimic a reddened photosphere.
The bulk of cluster and field stars in Fig.~\ref{vi-hk} lie at positions
consistent with normal photospheric colors; to the right of the dashed
line, the $H-K$ color is too red, however, for a normal reddened photosphere,
and there is an excess emission in the $K$ band, probably because of the
presence on a circumstellar dust disk. Tens of X-ray detected stars in
the PMS band (in the optical CMD) also fall in this part of the diagram,
consistent with their being CTTS; several of them are also \ha-excess
stars. Some care must be used, however, before claiming that these stars
do really possess an IR excess (and a dusty disk) because of their
placement in this diagram, because such a mixed representation is
obtained by matching two large catalogues, and spurious matches would
easily produce ``stars'' with strange colors. We have therefore
evaluated the number of spurious matches using the method explained in
Damiani \e (2006b), most appropriate when two matched catalogues contain
an expected large number of truly common objects, like here. The result
is a predicted number of 8.5 spurious matches in the ACIS FOV, between
2MASS and the SSB catalogue, or 0.23\% of total 2MASS-SSB matches in the
ACIS FOV (3685). This number is
much smaller than the number of stars to the right of the dashed line in
Fig.~\ref{vi-hk}, and therefore most of these latter must be true
IR-excess stars. However, we may take a step further, by considering the
match between the VPHAS$+$ and 2MASS catalogues. Although the SSB and
VPHAS$+$ photometry have most stars in common, they are statistically
independent datasets, with independent errors: therefore,
spurious matches between SSB and 2MASS will not generally coincide with those
between VPHAS$+$ and 2MASS. By matching these two latter catalogues, we
obtain the $(r-i,H-K)$ diagram of Figure~\ref{ri-hk}, which again shows
a large number of candidate IR-excess stars, with an estimated number of
16.7 spurious matches (0.38\% of all 4441 matches in the FOV).
If the two set of matches were completely independent,
the joint probability that an
optical-2MASS match is spurious in both the SSB and VPHAS$+$ cases is
$0.0023 \times 0.0038 = 8.74 \cdot 10^{-6}$, i.e.\ much less than one
star in the whole dataset; in reality, some covariance can be expected
so this estimate is excessively optimistic.
Nevertheless, the intersection between
the candidate IR-excess stars found in Figures~\ref{vi-hk} and~\ref{ri-hk}
may be considered
as a high-confidence IR-excess star sample (only these high-confidence
stars are indicated with orange triangles in the Figures).
Disregarding errors on the optical and IR colors, their
number would be 163 stars.
Taking instead a more conservative limit of a $3\sigma$ minimum distance from 
the fiducial limit, this number reduces to 48 IR-excess stars, which we
adopt here as the most reliable sample.
Had we taken the union of the two IR-excess samples,
their number would have been of 79 stars.
Only 7 of the high-confidence IR-excess stars are also \ha-emission stars.
We will examine the membership status of the IR-excess stars below.

In the most favorable case of 100\% membership, the number of IR-excess
stars would imply a circumstellar disk frequency of 4.6\% for NGC~6231 PMS
stars (becoming 15.7\% if the IR-excess star selection is performed less
conservatively, as often made).
This percentage is much smaller than that found in
clusters like the ONC (where almost all stars are found to have disks),
but compares well to other high-mass clusters such as NGC~2362 (1500~pc,
age 5~Myr),
where different studies converged to an estimated disk percentage of $\sim
10$\% (Haisch et al.\ 2001; Damiani et al.\ 2006a).

\begin{figure}
\resizebox{\hsize}{!}{
\includegraphics[bb=5 10 485 475]{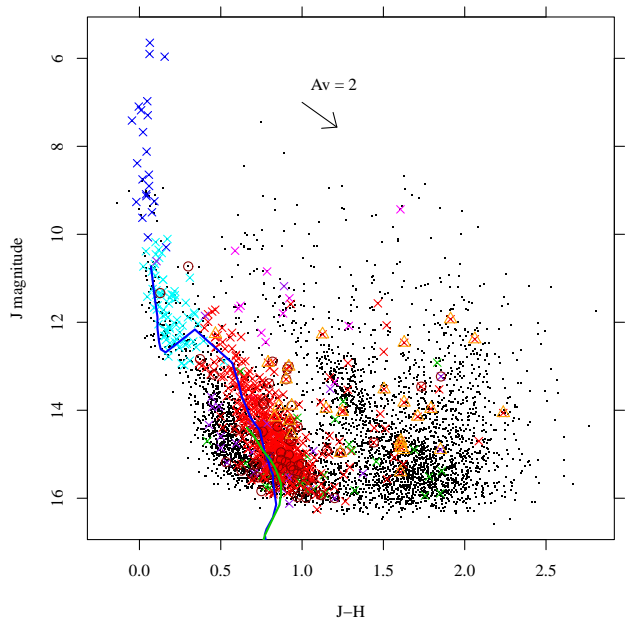}}
\caption{The 2MASS $(J,J-H)$ color-magnitude diagram for all IR sources
in the ACIS FOV. Symbols are as in Figure~\ref{v-vi-xray}.
The blue (green) line is a Siess (BHAC) isochrone at 5~Myr.
A representative reddening vector is also shown.
Symbols as in Fig.~\ref{sung-halpha}.
\label{j-jh-xray}}
\end{figure}

Figure~\ref{j-jh-xray} is a $(J,J-H)$
CMD for all 2MASS sources in the ACIS FOV. The X-ray selected stars
follow well the Siess and BHAC 5-Myr isochrones (the scatter being too
large to enable discrimination between the two). The X-ray undetected
field stars appear to split in two main groups: one at nearly the same
$J-H$ colors as the cluster, and a much redder one at $J-H \sim 1.5-2$.
Appendix~\ref{extinct-map} discusses the IR color distribution of field
stars in more detail.
Almost all the \ha-emission stars selected above fall in the
low-extinction group.
About one-half of the IR-excess stars defined above lie at low extinction,
while the other half are found in the high-extinction group, probably
because of local, circumstellar dust absorption.
A dozen of optically unidentified X-ray sources (green crosses)
are here found among the high-extinction 2MASS sources, free from IR
excesses: these are probably background, X-ray bright objects unrelated
to the cluster.

\begin{figure}
\resizebox{\hsize}{!}{
\includegraphics[bb=5 10 485 475]{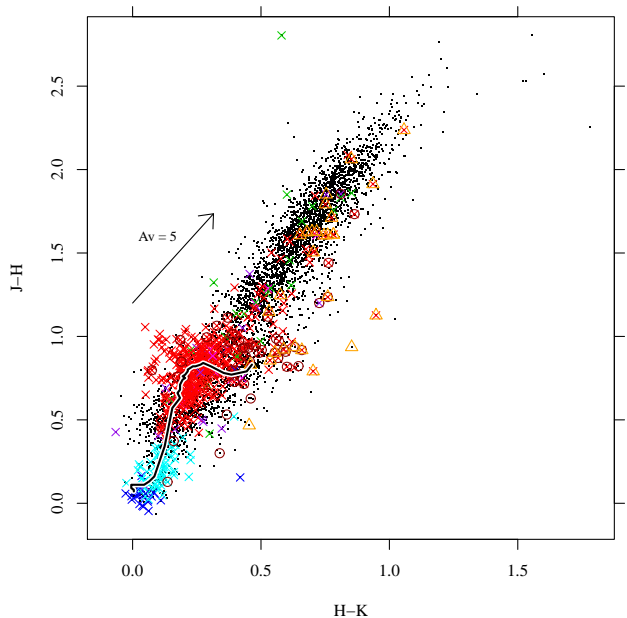}}
\caption{2MASS $(J-H,H-K)$ color-color diagram. Symbols are as in
Figure~\ref{v-vi-xray}.
\label{jh-hk-xray-zoom}}
\end{figure}

When considering the $(J-H,H-K)$ color-color diagram
(Figure~\ref{jh-hk-xray-zoom}), the X-ray detected and \ha-excess stars
are both found at a place consistent with low and uniform extinction.
Most of the high-extinction objects just discussed fall on a reddened-MS path.
In this diagram, less than a dozen stars exhibit significant
$K$-band excesses with respect to reddened MS colors,
among which only three detected in X-rays. A few tens additional stars
have possible excesses as well, but at a low significance level, and
are not considered. Note that in any case the population of X-ray
sources in the ``T~Tauri star locus'' (Meyer et al.\ 1997) is considerably
less numerous than in the IR color-color diagram of other massive clusters
like NGC6530 (Damiani \e
2006b) or the (less massive) ONC, suggesting that stars in NGC~6231
are found in a different (later) stage of disk-clearing with respect to
those younger clusters.
Also in the Cyg~OB2 association, more massive than NGC~6231, the 2MASS
color-color diagram yielded only a low number of IR-excess PMS stars
(Albacete-Colombo \e 2007).

Figure~\ref{jh-hk-xray-zoom} does not show an extremely red object
(2MASS J16532258-4148294: $J-H=4.59$, $H-K=3.34$) falling off to the
right, which at first seemed an embedded protostar (with a $K$ band excess),
none of which are known in NGC~6231 (we noted
above that no stars younger than about 2~Myr are found in the cluster).
This object falls in the North-Western outer part of the ACIS FOV.
From its $J-H$ and $H-K$ colors, an optical extinction of $A_V \sim
25$~mag is estimated; it has no optical nor X-ray counterpart in our data.
Rather than with a protostar, it is probably better identified with
the strong IR source IRAS~16498-4143
(within 4.6$^{\prime\prime}$) and with the maser source OH~343.4$+$1.3
(at 0.68$^{\prime\prime}$), which is a candidate OH/IR giant
(Sevenster et al.\ 1997).

\section{Spatial distributions of stellar groups}
\label{spatial}

\begin{figure}
\resizebox{\hsize}{!}{
\includegraphics[bb=5 10 485 475]{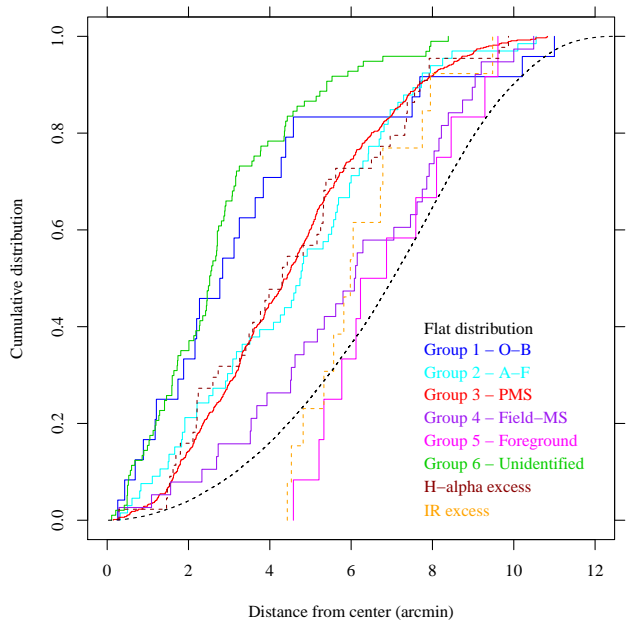}}
\caption{Cumulative distributions of distances from cluster center for
X-ray detected sources (with more than 20 counts) in different groups,
compared to a spatially uniform, flat distribution.
\label{profile-ir-xray-2}}
\end{figure}

\begin{figure}
\resizebox{\hsize}{!}{
\includegraphics[]{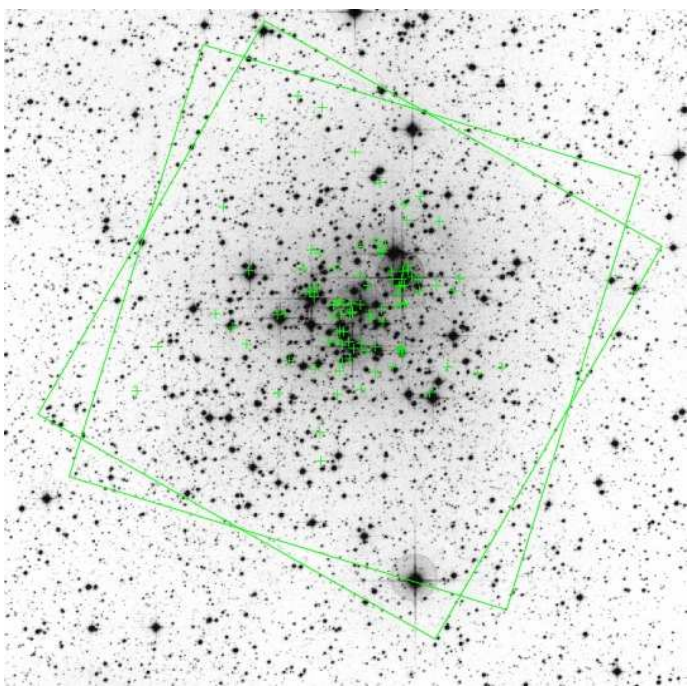}}
\caption{DSS2-red image of NGC~6231, with superimposed positions of
optically-unidentified X-ray sources (green crosses).
The boundaries of the two ACIS FOVs are also indicated.
\label{dss_unident}}
\end{figure}

\begin{figure}
\resizebox{\hsize}{!}{
\includegraphics[bb=5 10 485 475]{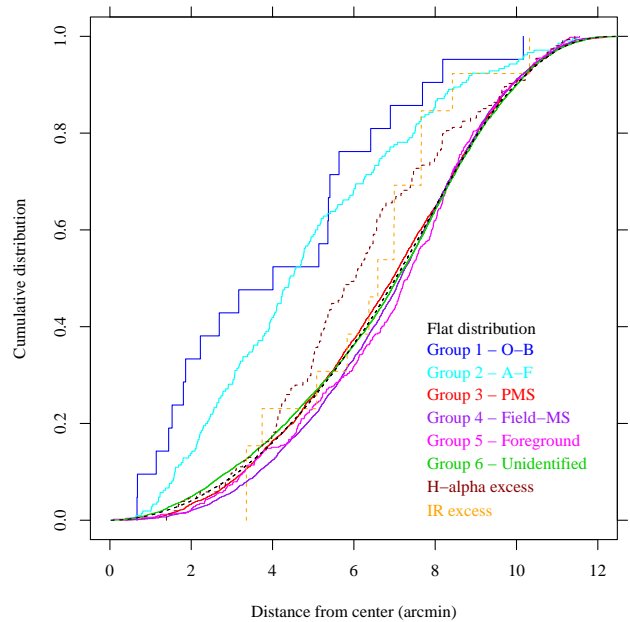}}
\caption{Cumulative distributions of distances from cluster center for
X-ray undetected sources in different groups.
\label{profile-ir-noxray-2}}
\end{figure}

Each of the groups defined above contains a large enough number
of stars that their spatial distributions carry significant information
on the group content.
We have computed the cumulative distributions of distances
from cluster center (defined in section~\ref{morph}) for X-ray
detected sources in all groups, as shown in Figure~\ref{profile-ir-xray-2}.
Since the X-ray detection threshold is not uniform across the
FOV but rises from 4-5 X-ray counts near field center to $\sim 20$
counts at the outer border, spatially-unbiased distributions require
consideration of detections with more than 20~counts only.
The expected distribution of objects falling uniformly across the
combined ACIS FOV (`flat distribution') is also shown for comparison. 
Groups 1-6 are disjoint samples (no common stars), while \ha-excess and
IR-excess samples are actually subsets of the other groups
(uncategorized by group).
Stars in groups 1-3 (OB, A-F, and PMS stars) are obviously more
strongly concentrated than the uniform distribution.
OB stars are much more concentrated than the A-F stars, which we take as
evidence of mass segregation, as already claimed for NGC~6231 massive
stars by Raboud and Mermilliod (1998), Raboud (1999), and SSB.
The same phenomenon was found in other massive clusters (e.g.\
$\lambda$ Ori, Dolan and Mathieu 2001, 2002; NGC~2362, Damiani \e 2006a).
Segregation seems instead not present among the X-ray detected A-F
stars, whose spatial distribution is nearly identical to that of PMS stars.

The very centrally concentrated distribution of the 
optically unidentified X-ray sources in group~6, very similar to that of
the OB stars, is very interesting. We rule out that these are OB stars,
since the SSB catalog is complete at bright magnitudes; they cannot even
be dominated by highly-reddened OB stars, since only few of them appear
as reddened objects in the IR diagrams discussed above, and the X-ray
spectral data (to be discussed below) indicate no particularly large
extinction for the bulk of them. Therefore, the best explanation for
their strong central clustering is that their missed optical
counterparts are faint cluster stars hidden in the glare of the
brightest cluster stars (10-15 magnitudes brighter).
This is supported by the DSS image in Fig.~\ref{dss_unident}, where we
show the positions of the unidentified X-ray sources (green crosses),
nearly coincident with the brightest stars.
The Chandra X-ray image of NGC~6231 does not suffer from glare, because the
luminosity contrast between the high-mass stars and the lowest-mass ones
is attenuated by a factor of $\sim 10^4$ with respect to the optical
(X-ray to bolometric luminosity ratios will be discussed below).
Therefore, a large fraction of the optically unidentified X-ray sources
should be considered as cluster members. We discuss
these stars in greater detail on the basis of the X-ray data below.

The X-ray detected \ha-emission stars (brown dashed line in
Fig.~\ref{profile-ir-xray-2}) have a spatial distribution
extremely similar to that of cluster PMS stars, and are indeed strongly
dominated by CTTS members of the cluster. The distribution of IR-excess
stars is markedly wider, instead, and more similar to a flat
distribution. This may be attributed to (at least) three causes: 1) the
IR-excess stars are generally not cluster members; 2) 2MASS data near
the cluster center are biased; 3) genuine cluster members with disk avoid the
central cluster region and the OB stars concentrated there.
Of these hypotheses, the first seems unlikely, since the IR-excess stars
found here are too many for a field population, at a distance no greater
than the cluster.
The second hypothesis is partially
true, since the spatial density of 2MASS sources fainter than $J \sim
15$ is found to be lower
near cluster center than in the outer parts, which is again explained by
the glare of massive stars preventing detection of the faintest stars nearby; 
we checked, however, that the difference between the spatial distributions of
PMS stars in group~3 and IR-excess stars still remains for brighter 2MASS
sources ($J<15$), at a significance level of 99.67\%: therefore, this
favors our third hypothesis, that indeed circumstellar disks tend to
avoid the vicinity of OB stars, where they are disrupted on shorter
timescales. A similar result was obtained 
by Guarcello \e (2007, 2010b) in the case of the massive cluster NGC~6611,
and very recently by Zeidler \e (2016) for Westerlund~2.
X-ray detected stars in the Field-MS and Foreground groups have spatial
distributions similar to a flat distribution, consistently with these
groups being composed of field stars.
The statistical significance of the pairwise differences between
these distributions was tested by means of Kolmogorov-Smirnov tests,
whose results are reported in Table~\ref{ks-table}.

\begin{center}
\begin{table*}[ht]
\centering
\caption{K-S test probabilities (\%) of no difference between subgroups} 
\label{ks-table}
\begin{tabular}{lrrrrrrrr}
  \hline
 & A-F & PMS & Field-MS & Foreground & Unidentified & \ha\ excess & IR excess & Flat \\ 
  \hline
O-B & 1.40 & 2.90 & 0.03 & $<$0.01 & 72.00 & 11.00 & $<$0.01 & $<$0.01 \\ 
  A-F &  & 54.00 & 4.30 & 1.30 & $<$0.01 & 77.00 & 1.80 & $<$0.01 \\ 
  PMS &  &  & 0.69 & 0.04 & $<$0.01 & 88.00 & 0.33 & $<$0.01 \\ 
  Field-MS &  &  &  & 26.00 & $<$0.01 & 6.90 & 51.00 & 14.00 \\ 
  Foreground &  &  &  &  & $<$0.01 & 0.42 & 59.00 & 59.00 \\ 
  Unidentified &  &  &  &  &  & 0.01 & $<$0.01 & $<$0.01 \\ 
  \ha\ excess &  &  &  &  &  &  & 0.83 & $<$0.01 \\ 
  IR excess &  &  &  &  &  &  &  & 17.00 \\ 
   \hline
\end{tabular}
\end{table*}
\end{center}

Figure~\ref{profile-ir-noxray-2} shows instead the spatial distributions
of stars undetected in X-rays (or detected with less than 20 counts)
in the various groups, in the ACIS FOV. Here stars in groups 3-6 have
distributions nearly identical to a flat distribution.
The X-ray undetected OB stars have instead a centrally-clustered
distribution, but less so than the X-ray detected OB stars of
Fig.~\ref{profile-ir-xray-2}. This intermediate degree of central
clustering may be explained by noting that the X-ray undetected massive
stars are in fact all of B type (Fig.~\ref{v-bv-xray-himass}), and these
may be less centrally segregated than the more massive stars. Because of
their rarity, both X-ray detected and undetected OB stars must be
cluster members.
The X-ray undetected A-F stars also have a spatial distribution
similar to the undetected B stars, and nearly identical to that of the X-ray
detected A-F stars of Fig.~\ref{profile-ir-xray-2}. Therefore, the
fraction of field stars in both the X-ray detected and undetected A-F
groups must be very nearly the same. The X-ray detected
A-F stars have in turn a distribution almost coincident with the cluster PMS
stars, as mentioned above, which suggests that (nearly) all A-F stars,
either X-ray detected or not, are cluster members.
To avoid this conclusion, one should hypothesize a rather ad-hoc
combination of mildly-segregated A/B-type stars plus some number of
uniformly-distributed A-F field stars, whose combined spatial
distribution would mimic that of cluster PMS stars: this seems
rather unlikely.

The distribution of \ha-emitting stars not detected in X-rays is
intermediate between a flat one and the distribution of cluster PMS
stars. The \ha\ thresholds used to define the \ha-emitting sample are
appropriate to CTTS stars (above the chromospheric \ha\ level), which
are not expected to be found among field stars. If all \ha-emitting
stars are indeed members, such broader distribution with respect to all
cluster PMS stars might again be a clue that accretion from
circumstellar disks terminates earlier in the innermost cluster regions
where OB stars are found. The distribution of X-ray undetected IR-excess
stars is very similar to that of X-ray detected ones, so again the same
considerations apply.

\section{X-ray spectra}
\label{xspec}

The low-resolution X-ray spectra provided by ACIS are useful both to
help classifying the detected sources, and to obtain information on the
nature of their X-ray emission. Most of the X-ray sources are relatively
faint: only 16 have more than 500 X-ray counts (all massive stars or
flaring Group~3 stars), and 176 have more than 100
counts. Therefore, only for a very small fraction of sources we can
attemp to fit their individual spectra to obtain meaningful spectral
parameters. However, some spectral information can also be obtained by
considering X-ray hardness ratios, even for sources with few
counts.

\begin{figure}
\resizebox{\hsize}{!}{
\includegraphics[bb=5 10 485 475]{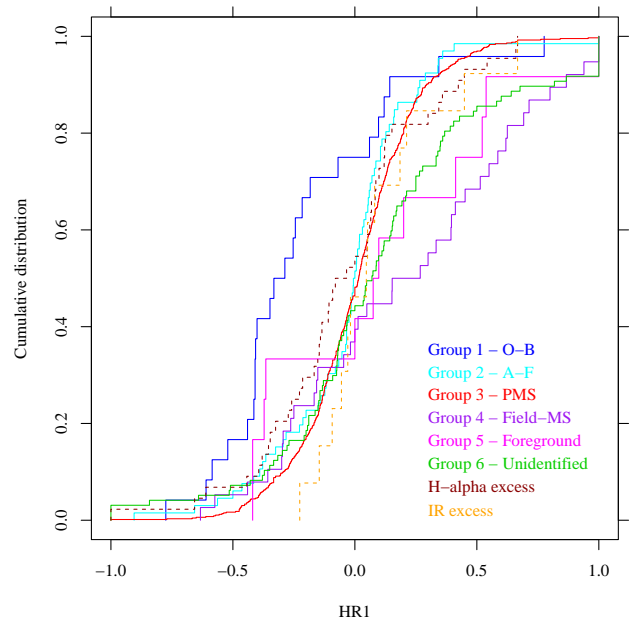}}
\caption{Cumulative distributions of hardness ratio HR1 for non-flaring stars
in different groups.
\label{hr1-distr-ir-groups}}
\end{figure}

\begin{figure}
\resizebox{\hsize}{!}{
\includegraphics[bb=5 10 485 475]{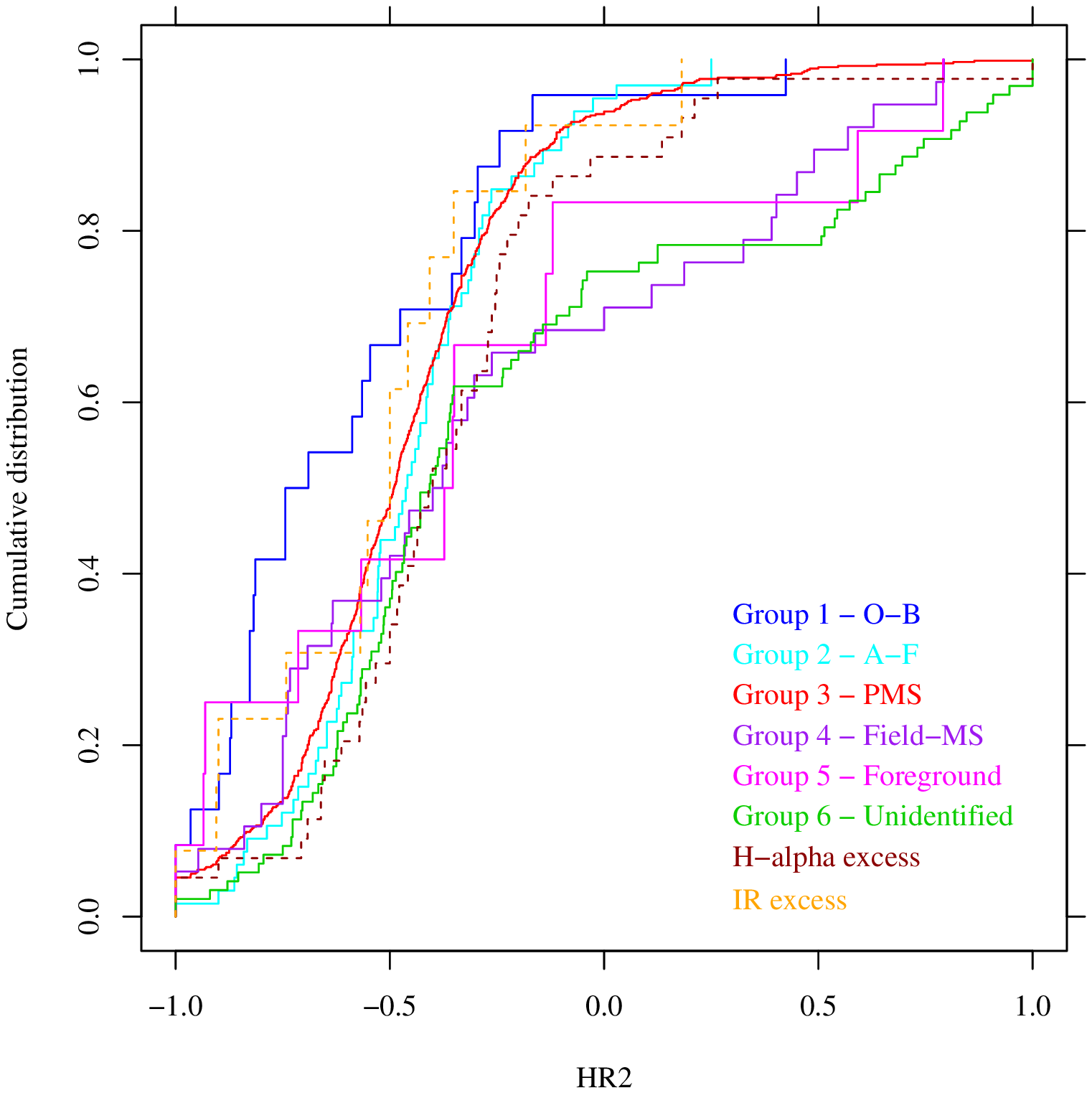}}
\caption{Cumulative distributions of hardness ratio HR2 for non-flaring stars
in different groups.
\label{hr2-distr-ir-groups}}
\end{figure}

\begin{figure}
\resizebox{\hsize}{!}{
\includegraphics[bb=5 10 485 475]{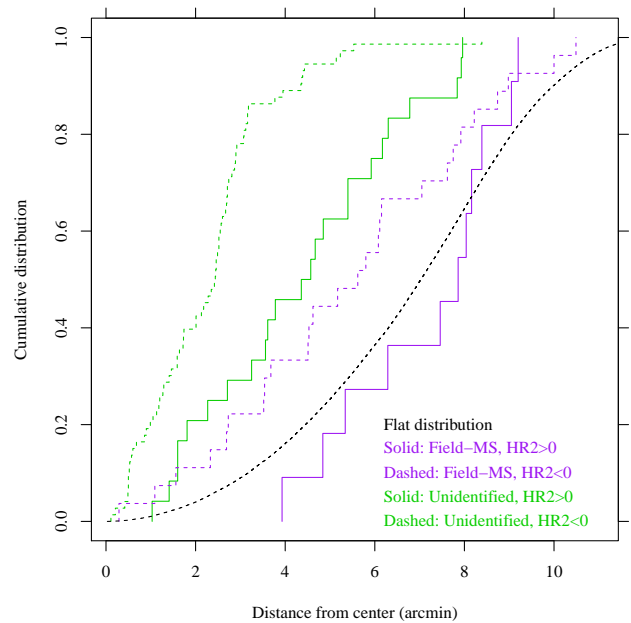}}
\caption{Cumulative distributions of distances from cluster center for
X-ray detected sources in groups 4 (Field-MS) and 6 (unidentified),
separating hard (HR2$>0$) and soft (HR2$\leq 0$) sources, respectively.
The dotted line is a spatially uniform distribution.
\label{profile-ir-soft-hard}}
\end{figure}

\subsection{X-ray hardness ratios}
\label{hr}

We define here two hardness ratios as $HR1 = (M-S)/(M+S)$, and 
$HR2 = (H-M)/(H+M)$, where $S$, $M$, and $H$ are the source X-ray counts
in the soft (0.3-1.2 keV), medium (1.2-2.2 keV), and hard (2.2-8.0 keV)
bands, respectively.
Relatively high values of HR1 and HR2 are indicative of
harder (hotter) X-ray emission, while a high HR1 is also indicative of high
absorption, the latter hiding selectively the softer emission.
The hardness ratios of our detections are reported in Table~\ref{table1}.
We show the cumulative distributions of HR1 and HR2 for the different
groups of sources in NGC~6231 in Figures~\ref{hr1-distr-ir-groups}
and~\ref{hr2-distr-ir-groups}, respectively. Only non-flaring sources
with more than 20 X-ray counts
are included in these distributions, since flaring spectra may be much
harder than quiescent ones, and introduce a bias\footnote{Flaring sources
are here selected as having lightcurves statistically incompatible
with both a uniform level, and a constant slope.}.
Figure~\ref{hr1-distr-ir-groups} shows that the median HR1 is very
nearly the same for all groups, except for the OB stars,
which have a lower median HR1. We have discussed in Section~\ref{cmd}
that OB and low-mass PMS stars share a common
average extinction value, and therefore the low median HR1 for OB
stars is unlikely to be related to a lower absorption. Since massive
stars are typically softer X-ray sources than lower-mass stars, the
lower HR1 is more probably related to a predominantly softer emission
rather than to an increased absorption with respect to lower-mass X-ray
sources. Figure~\ref{hr1-distr-ir-groups} shows also some difference
among the different groups in the upper tail of the HR1 distribution,
with an extended tail more prominent among Field-MS and unidentified X-ray
sources.  This is still more obvious from the HR2 distribution of
Figure~\ref{hr2-distr-ir-groups} (where we find again a lower median for
OB stars, related to their soft emission spectrum): here sources in the
Field-MS and unidentified groups are distributed in a distinctly different
way than other sources, showing enough X-ray emission above 2.2 keV to
raise the HR2 value above zero in $\sim$35\% of cases.

The discussion above suggests that the majority of the Field-MS sources are
not cluster members, and we will not discuss them further; on the other
hand, we argued that most of the unidentified X-ray sources are good
candidate members, so that their HR2 distribution being so different
from that of the cluster stars is puzzling.
Fig.~\ref{hr2-distr-ir-groups} suggests that a threshold near HR2=0 may
be an effective means of separating X-ray sources of different nature
inside these two groups. Accordingly, the spatial distributions of
sources in these two groups are shown again in
Figure~\ref{profile-ir-soft-hard}, but separately for sources having HR2
less and greater than zero, respectively.
The Figure shows that the softer sources in each group have a more
centrally concentrated distribution than the hard sources. In
particular, this reinforces our earlier suggestion that the unidentified
sources with HR2$<0$ are all cluster members. The limited statistics of
the two Field-MS subsamples is such as the distributions of both the harder
and softer sources are compatible with a flat distribution
(non-members). A bit puzzling is the distribution of unidentified
sources with HR2$>0$, found significantly different than a flat
distribution (99.98\% level), which would suggest that these unusually
hard (but non-flaring), weak X-ray sources are clustered towards the center of
NGC~6231. A caveat against this conclusion might be that unidentified
X-ray sources are rarer in the outer part of the FOV (regardless of
their spectrum) simply because the wider ACIS PSF there ensures more
counterparts to be found (including spurious ones).
Therefore, we conclude that only the unidentified X-ray sources with
HR2$<0$ should be reliably included among cluster candidate members.

\subsection{X-ray spectral fits}
\label{xspecfit}

\begin{table*}[ht]
\centering
\caption{X-ray fitting results for combined spectra. Emission-measure
units are only relative. Conversion factors have units of 10$^{-11}$ erg
cm$^{-2}$ count$^{-1}$.} 
\label{table-xmod}
\begin{tabular}{lrrrrrrrrrr}
  \hline
Group & No. & Total & $N_H$ & $kT_1$ & $EM_1$ & $kT_2$ & $EM_2$ &
$kT_3$ & $EM_3$ & Conv.\ \\ 
& src.\ & counts & ($10^{22}$ cm$^{-2}$) & (keV) & & (keV) & &
(keV) & & fact.\ \\ 
  \hline
OB & 18 & 2702 & 3.31E-01 & 0.343 & 4.61E-01 & 0.989 & 1.54E-01 & 3.37 & 6.85E-02 & 2.50 \\ 
A-F & 86 & 5305 & 5.41E-01 & 0.354 & 5.42E-01 & 1.08 & 4.57E-01 & 2.69 & 2.16E-01 & 2.62 \\ 
PMS-bright & 345 & 27666 & 4.25E-01 & 0.432 & 1.03E+00 & 1.04 & 1.69E+00 & 2.49 & 1.49E+00 & 1.95 \\ 
  PMS-faint & 428 & 14004 & 2.64E-01 & 1.06 & 5.72E-01 & 2.73 & 8.65E-01 & 2.73 & 3.03E-02 & 1.46 \\ 
  Field-MS & 75 & 1455 & 1.13E+00 & 0.2 & 6.03E+00 & 3.09 & 2.10E-01 & ... & ... & 26.4 \\ 
  Foreground & 15 & 719 & 0.33E+00 & 0.785 & 4.09E-02 & 3.32 & 4.66E-02 &
... & ... & 1.83 \\ 
  Unident.-soft & 226 & 4796 & 3.80E-01 & 0.2 & 3.96E-01 & 0.992 & 2.65E-01 & 2.08 & 2.96E-01 & 2.01 \\ 
  IR-excess & 21 & 1082 & 7.68E-01 & 0.2 & 2.08E+00 & 1.78 & 1.23E-01 & 15.0 & 1.22E-02 & 12.9 \\
  H$\alpha$ & 54 & 2419 & 6.79E-01 & 0.2 & 2.20E+00 & 0.343 & 7.84E-02 & 1.55 & 3.24E-01 & 7.08 \\ 
   \hline
Group & No. & Total & $N_{H1}$ & $kT_1$ & $EM_1$ & $N_{H2}$ & $kT_2$ & $EM_2$ &
& Conv.\ \\ 
& src.\ & counts & ($10^{22}$ cm$^{-2}$) & (keV) & & ($10^{22}$ cm$^{-2}$) &
(keV) & & & fact.\ \\ 
  \hline
  Unident.-hard & 100 & 1901 & 4.76 & 1.35 & 3.61E-02 & 3.93 & 6.34 &
4.42E-01 & ... & 4.61 \\ 
   \hline
\end{tabular}
\end{table*}

\begin{figure}
\resizebox{\hsize}{!}{
\includegraphics[bb=5 10 485 475]{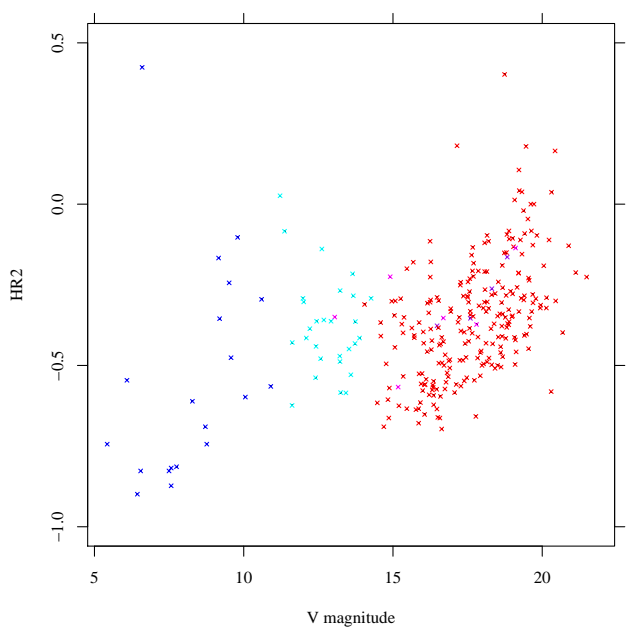}}
\caption{Hardness ratio HR2 vs.\ $V$ magnitude for OB, A-F, and PMS stars.
Symbols as in Fig.~\ref{v-vi-xray}.
\label{vmag-hr2}}
\end{figure}

\begin{figure*}
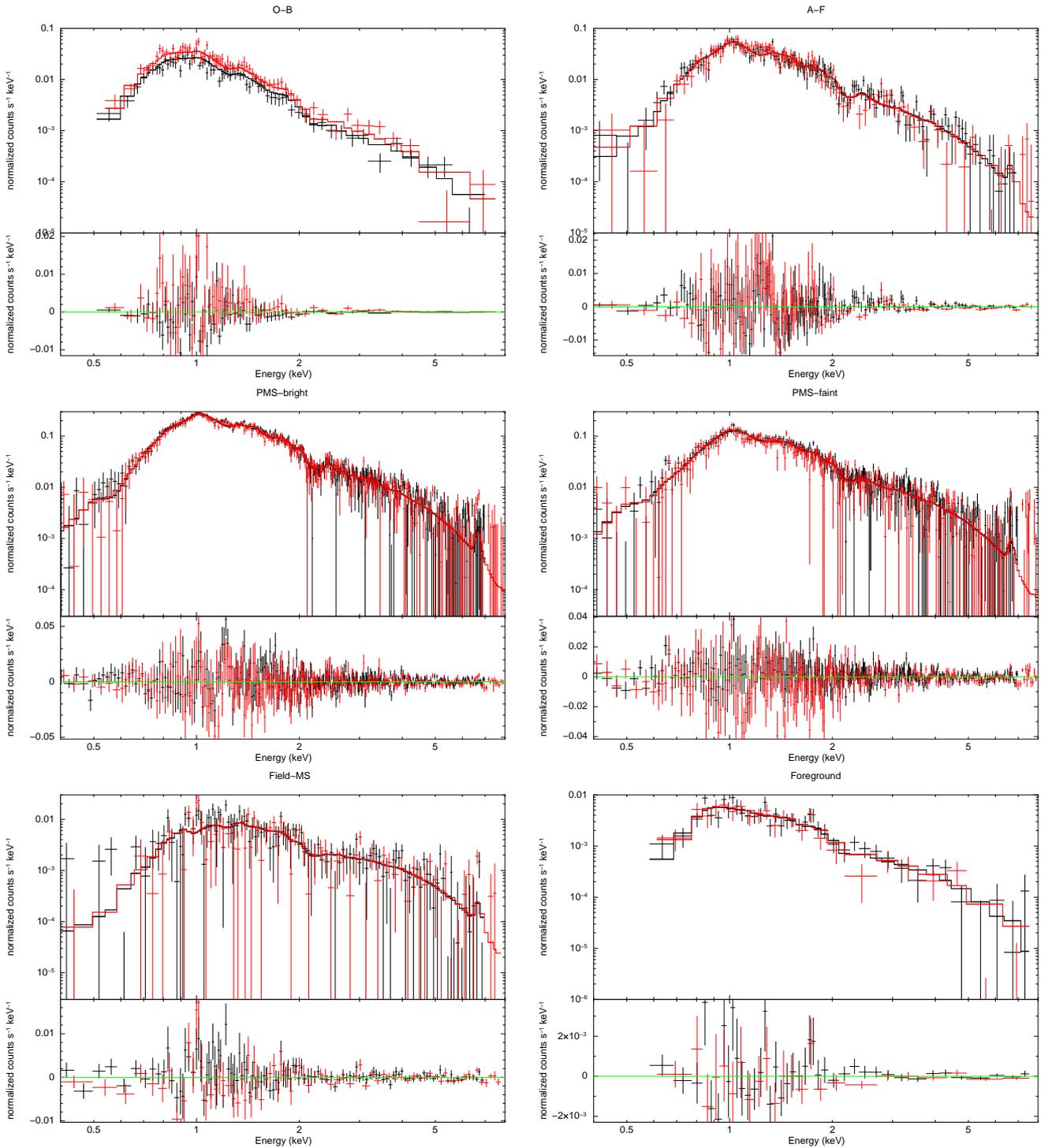

\centering
\includegraphics[angle=-90,width=88mm]{combin_1_3T_NHfree_kTfree_abu0.3_DEMfree.eps}
\includegraphics[angle=-90,width=88mm]{combin_2_3T_NHfree_kTfree_abu0.3_DEMfree.eps}
\includegraphics[angle=-90,width=88mm]{combin_3_3T_NHfree_kTfree_abu0.3_DEMfree.eps}
\includegraphics[angle=-90,width=88mm]{combin_4_3T_NHfree_kTfree_abu0.3_DEMfree.eps}
\includegraphics[angle=-90,width=88mm]{combin_5_3T_NHfree_kTfree_abu0.3_DEMfree.eps}
\includegraphics[angle=-90,width=88mm]{combin_6_2T_NHfix.eps}
\caption{Summed X-ray spectra (crosses) of sources in groups 1-5,
with model fits. The black and red observed spectra
correspond to ObsIds 5372 and 6291, respectively.
\label{xspectra}}
\end{figure*}

\begin{figure*}
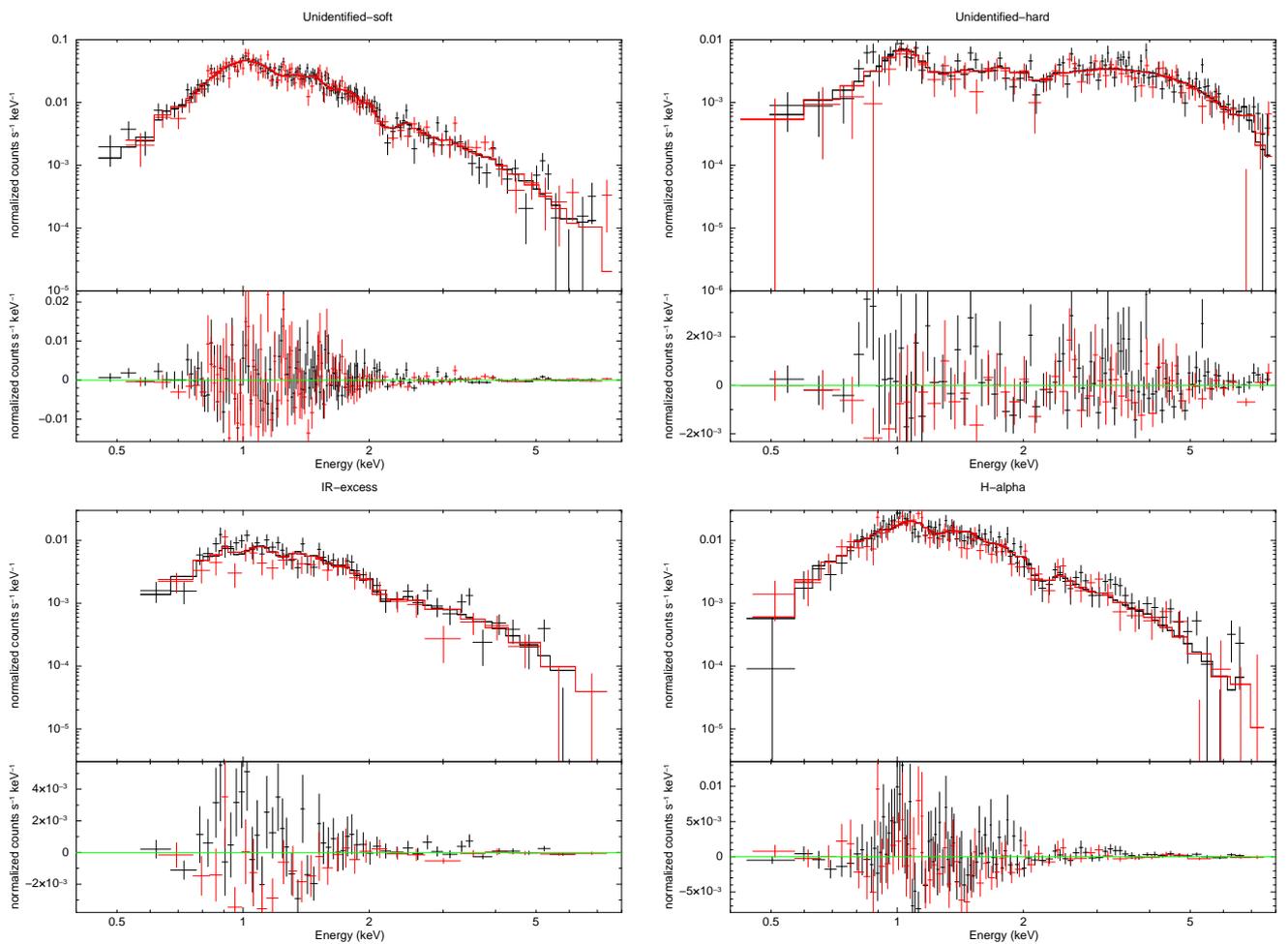

\centering
\includegraphics[angle=-90,width=88mm]{combin_7_3T_NHfree_kTfree_abu0.3_DEMfree.eps}
\includegraphics[angle=-90,width=88mm]{combin_10_2T-2NH.eps}
\includegraphics[angle=-90,width=88mm]{combin_8_3T_NHfree_kTfree_abu0.3_DEMfree.eps}
\includegraphics[angle=-90,width=88mm]{combin_9_3T_NHfree_kTfree_abu0.3_DEMfree.eps}
\caption{Summed X-ray spectra of optically unidentified sources with
$HR2<0$ (top left) and $HR2>0$ (top right), and of 
\ha-excess and IR-excess stars (as labeled), with model fits.
\label{xspectra2}}
\end{figure*}

As said above, only very few X-ray sources in NGC~6231 are detected with more
than 500
counts, and therefore X-ray spectral parameters cannot be reliably determined
for the great majority of sources. However, the cumulative number of X-ray
counts for all sources in a given group is generally high enough for a
meaningful spectral fit, assuming that their intrinsic spectra are
homogeneous enough for this treatment to make sense.
The above distributions of hardness ratios HR1, HR2 are narrow enough to
support this assumption, except for the hard/soft dichotomy among the
unidentified sources, which were therefore treated separately.
While in principle also the Field-MS sources would need to be similarly splitted
into a soft and hard group, in practice their total X-ray counts are too
few to permit this.
Moreover, the richest group of PMS stars was split in two subgroups,
respectively bluer and redder than $V-I=2.2$.
This was motivated by the apparent trend of increasing HR2 towards fainter
PMS members, as shown in Figure~\ref{vmag-hr2}.
Therefore, we proceeded
by computing summed spectra of
all sources in a group and fitting the average spectra thus obtained,
using Xspec v12. We
excluded from this procedure sources showing flares, and also the 9
non-flaring, massive sources with more than 500 counts\footnote{These
are the X-ray sources \#201, 377, 430, 647, 699, 1008, 1016, 1225, and
1379, identified with HD~152219, HD~152234, HD~152233, V1007~Sco=HD~152248,
HD~152249, HD~152270, V1034~Sco=CPD-41~7742, HD~326331, and HD~152314,
respectively.},
which are exceedingly stronger than the rest of stacked spectra, and
whose inclusion would make the group too inhomogeneous.
We jointly fit X-ray spectra from both ObsId 5372 and 6291; however, the
cumulative group spectrum from each ObsId has its own instrument
response function, and their combination is done only at the fitting stage
by Xspec.
We have attempted to fit these combined spectra using six different models,
choosing afterwards the one giving the most successful fits.
All six models are three-temperature (3T) APEC models.
In models~1-5,
the three temperatures are fixed at the values of 0.5, 1.0 and 2.0 keV,
respectively, while in model~6 they are allowed to vary.
The metal abundance is held fixed at 0.1 times solar in models 1-3, and at
0.3 solar in models 4-6.
Absorption $N_H$ is held fixed at $3.22 \times 10^{21}$ cm$^{-2}$
(as derived from the optical extinction) in models 2,3 and 5, and left free
in the other cases.
The emission-measure ratios between the three thermal
components are held fixed in models 1-2, at values 
$EM_{middle}/EM_{cool}=0.85$ and $EM_{hot}/EM_{cool}=0.7$, which gave good
results in the case of NGC~2362 (Damiani \e 2006a).
The first two models gave generally poorer fits than
models 3-6. The latter four models gave similar acceptance probabilities
for each spectrum; however,
visual inspection of observed spectra and model fits showed that some
features were consistently better reproduced by the last model, which has
the largest number of free parameters. We therefore adopt this latter
spectral model, for all groups
except for Field-MS and Foreground, which because of the small number of
total X-ray counts were fitted using a 2T model
(with fixed absorption $N_H$ in the case of Foreground sources).
Observed (cumulative) spectra and model fits are shown
in Figures~\ref{xspectra} and~\ref{xspectra2}.

For unidentified hard sources the 3T model with a single absorption
value yielded always unsatisfactory fits. A much better model had two
free temperatures, each component with a different absorption $N_H$,
left free to vary (metal abundance were instead set to 0.3 times
the solar value).
The fit yielded
a $\chi^2/d.o.f. = 173.9/144$ (probability $P=0.045$), and best-fit
parameters $kT_1=1.35$~keV, $N_{H1}=4.76 \cdot 10^{20}$~cm$^{-2}$ for the
soft component, and $kT_2=6.34$~keV, $N_{H2}=3.93 \cdot 10^{22}$~cm$^{-2}$
for the hard component.
Table~\ref{table-xmod} reports the best-fit parameters for each group. Also
listed is the count-rate to flux conversion factor, corrected for
absorption, which is used to compute source X-ray luminosities (in
the 0.4-8.0 keV band), listed in Table~\ref{table1}.
In particular, the conversion factor for Group~3 sources, assuming a
minimum number of 20 X-ray counts for a detection anywhere in the FOV,
implies for the PMS members completeness down to 
$L_X \sim 8.4 \times 10^{29}$ erg/s.
The minimum detected X-ray luminosity for
optically unidentified soft sources (and likely members as well) is
$L_X \sim 2.0-2.4 \times 10^{29}$ erg/s, attained only in the FOV center.

The best-fit absorption $N_H$ from Table~\ref{table-xmod} for OB, A-F,
and PMS stars is consistent with the value inferred from optical
extinction. IR- and \ha-excess stars have best-fit $N_H$ twice as large,
which might indicate higher gas and dust column densities towards these
stars; however, the number of X-ray counts used in the fits are smaller
than for the other member groups, so the actual significance of the
increased best-fit absorption is unclear. The A-F stars have
X-ray temperatures very close to the PMS-bright stars, in agreement with
the hypothesis that their emission originates from coronae of colder
binary companions (which we discuss in detail below). The comparison
between best-fit temperatures of bright and faint PMS members confirms
that the average X-ray emission hardens towards the lower-mass stars,
even disregarding flaring stars. The best-fit temperatures of soft
unidentified sources (also to be discussed below) are approximately
consistent with those of the fainter PMS stars.

\section{X-rays from massive stars}
\label{himass}

\begin{table}[t]
\centering
\caption{Upper limits to X-ray luminosities of undetected massive stars.} 
\label{table-upp}
\begin{tabular}{rlllr}
  \hline
SBL & Second & Third & Spectral & lim.log $L_X$ \\ 
 & Ident. & Ident. & Type & (erg/s) \\ 
  \hline
 20 & HW 19 & HD 326326 & B2V & 30.50 \\ 
  113 & Se 28 & HD 326327 & B1.5IVe+shell & 30.09 \\ 
  134 & Se 30 &  & B7V & 30.02 \\ 
  157 & Se 34 & HD 326328 & B1V & 29.93 \\ 
  164 & Se 33 &  & B7V & 29.92 \\ 
  175 & Se 14 & NSV 20801 & B4IV & 30.10 \\ 
  262 & Se255 &  & B6V & 29.86 \\ 
  268 & Se282 &  & B2V+B2V & 29.72 \\ 
  272 & Se284 &  & B4.5V & 29.83 \\ 
  275 & Se249 &  & B4.5V & 29.84 \\ 
  301 & Se276 &  & B8Vp & 30.48 \\ 
  323 & Se194 &  & B3.5V & 29.99 \\ 
  340 & Se274 &  & B3V & 29.66 \\ 
  346 & SS 62 & CPD-41 772 & B1V & 30.90 \\ 
  349 & Se295 &  & B1V & 29.64 \\ 
  351 & SS 56 &  & B1V & 29.81 \\ 
  364 & Se 6 & CD-41 11031 & B0V & 29.97 \\ 
  374 & Se70 & HD 326340 & B0.5V & 31.03 \\ 
  378 & Se286 &  & B0.5V & 29.69 \\ 
  403 & Se287 & CD-41 1103 & B1V & 29.63 \\ 
  410 & Se243 &  & B8.5V & 29.71 \\ 
  422 & Se 95 &  & B8.5V & 30.06 \\ 
  437 & Se 80 & V963 Sco & B0Vn & 29.91 \\ 
  447 & Se 78 &  & B8.5III & 30.07 \\ 
  452 & Se235 &  & B8.5V & 29.62 \\ 
  464 & Se189 &  & B8.5V & 29.89 \\ 
  476 & Se234 &  & B8Vn & 29.62 \\ 
  480 & Se213 &  & B2IVn & 29.67 \\ 
  486 & Se238 & HD 326330 & B1V(n) & 29.65 \\ 
  497 & Se 73 & HD 326339 & B0.5III & 30.21 \\ 
  513 & Se227 &  & B7Vn & 29.65 \\ 
  515 & SS129 &  & B1Vn & 29.81 \\ 
  521 & Se232 & CPD-41 774 & B0.5V & 29.65 \\ 
  528 & Se223 &  & B6Vn & 29.66 \\ 
  529 & Se217 &  & B6Vn & 29.66 \\ 
  547 & Se108 &  & B2V & 29.78 \\ 
  572 & Se184 &  & B8.5V & 29.88 \\ 
  587 & Se222 &  & B4Vn & 29.71 \\ 
  643 & Se115 &  & B3V & 30.10 \\ 
  648 & Se152 &  & B5V & 29.99 \\ 
  651 & SS165 & HD 326332 & B1III* & 29.98 \\ 
  702 & Se173 &  & B9V & 30.11 \\ 
  712 & Se150 & HD 326333 & B1V(n) & 29.97 \\ 
  717 & Se123 &  & B6.5V & 30.21 \\ 
  726 &  &  & B9V & 30.20 \\ 
  773 & Se147 &  & B9.5V & 30.18 \\ 
   \hline
\end{tabular}
\end{table}

We have studied the X-ray emission of massive O and B stars in
the cluster. While they has been studied using XMM-Newton data by Sana
et al.\ (2006b), our higher-resolution Chandra image enables us to test if
their results are affected by source confusion, which may easily happen
in such a dense cluster. This may be especially true for the B stars, which
generally have X-ray luminosities comparable to lower-mass stars,
and are outnumbered by these latter
stars
by a large factor. We have therefore
computed $L_X$ upper limits for all massive stars listed by Sana et al.\
(2006b) which remained undetected in our observation, as
reported in Table~\ref{table-upp}. However, it turns out that all massive stars
positively detected in X-rays by Sana et al.\ (2006b) (their Table~2) are
also detected in our data (if falling within the ACIS FOV).
Their conclusions on X-ray emission from massive stars are therefore not
affected by source misidentification.
Since the XMM-Newton X-ray spectra of these stars presented
by Sana et al.\ (2006b) have a considerably larger number of photons
than the ACIS spectra in our datasets, we do not consider them further
here.

\begin{figure}
\includegraphics[angle=-90,width=78mm]{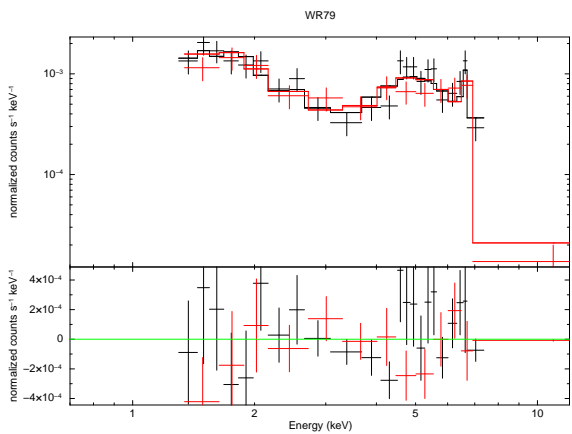}
\caption{The hard, peculiar X-ray spectrum of source \#1008 (HD~152270, WR~79),
with best-fit 2T model.
\label{xfit1008}}
\end{figure}

We have studied source \#1008 (the Wolf-Rayet star HD~152270 =
WR~79\footnote{This star must not be confused with WR~79a, lying more
than half a degree away outside the ACIS FOV, and studied
in X-rays by Skinner et al.\ (2010).}),
which was not studied by Sana et al.\ (2006b), and has a peculiar, hard
spectrum (Figure~\ref{xfit1008}).
Despite being clearly identified with
a massive star, this spectrum shares some similarities with that of the
fainter unidentified hard sources fitted above (hard and soft components with 
differing absorption), and is accordingly well fitted
($\chi^2$/d.o.f.\ =31.2/31) by a two-temperature
model ($kT = 1.1$ and 1.9~keV) with very different absorption ($N_H = 2.18
\cdot 10^{22}$ and 49.4 $\cdot 10^{22}$ cm$^{-2}$, respectively, the
latter being poorly constrained).
Compared to the above fits of unidentified hard sources,
the absorption of both components in WR~79 is much larger. It would
correspond (using an interstellar gas/dust ratio) to optical extinctions
towards the two components of 10 and $>$200 mag, respectively, far higher
than the star optical extinction (no different from the rest of the
cluster, judging from the optical two-color diagram). Examining the
X-ray image, the hard and soft components do not appear to be spatially
distinct, making unlikely a chance superposition with a heavily obscured
background source. Also the agreement between X-ray and optical/IR
positions is very good (0.17$^{\prime\prime}$ using the 2MASS position),
and the intensity (count rate) of the source is much larger than all
hard unidentified sources (where most background objects are expected to fall).
Therefore the observed spectrum is most likely intrinsic to WR~79.
The hard, strongly absorbed spectrum of WR~79 is reminiscent of that of
\object{WR~20a} in Westerlund~2 (Naz\'e, Rauw and Manfroid 2008), but
is still hotter and more absorbed than it; if interpreted in the same
framework, it suggests production of hard X-rays in stronger shocks
within a denser wind-wind collision zone, compared to WR~20a.
In the same way, the hard-component absorption is significantly higher
in WR~79 than in the WR stars in Westerlund~1 (Skinner et al.\ 2006a),
and in the colliding-wind binary \object{WR~147} (Skinner et al.\ 2007,
Zhekov 2007).
WR~79 is actually one of the few WR stars of type WC to be detected in
X-rays (Oskinova et al.\ 2003, Skinner et al.\ 2006b),
probably because of its binary nature.

\section{X-rays and stellar properties}
\label{xlumfn}

\begin{figure}
\resizebox{\hsize}{!}{
\includegraphics[bb=5 10 485 475]{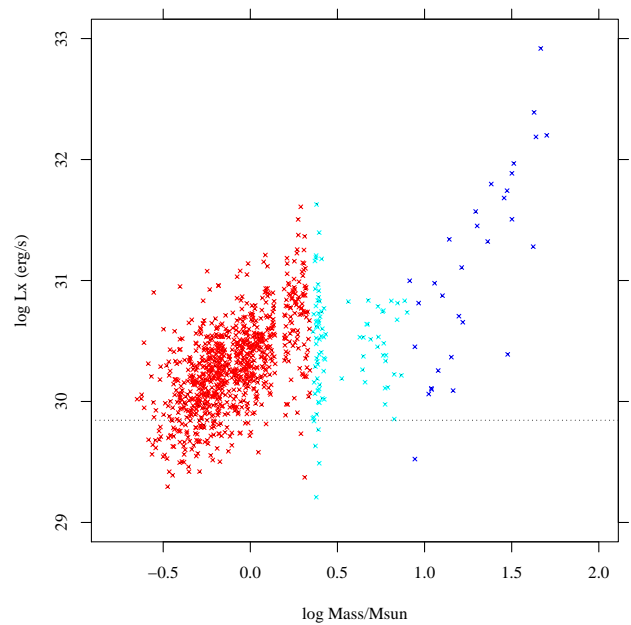}}
\caption{Plot of X-ray luminosity $L_X$ vs.\ stellar mass.
Symbols as in Fig.~\ref{v-vi-xray}.
The dotted horizontal line is our completeness limit.
The narrow gap at $\sim 1.4 M_{\odot}$ is an artifact corresponding to
the transition between the BHAC and Siess \e (2000) isochrone sets used.
\label{lx-mass-ir}}
\end{figure}

\begin{figure}
\resizebox{\hsize}{!}{
\includegraphics[bb=5 10 485 475]{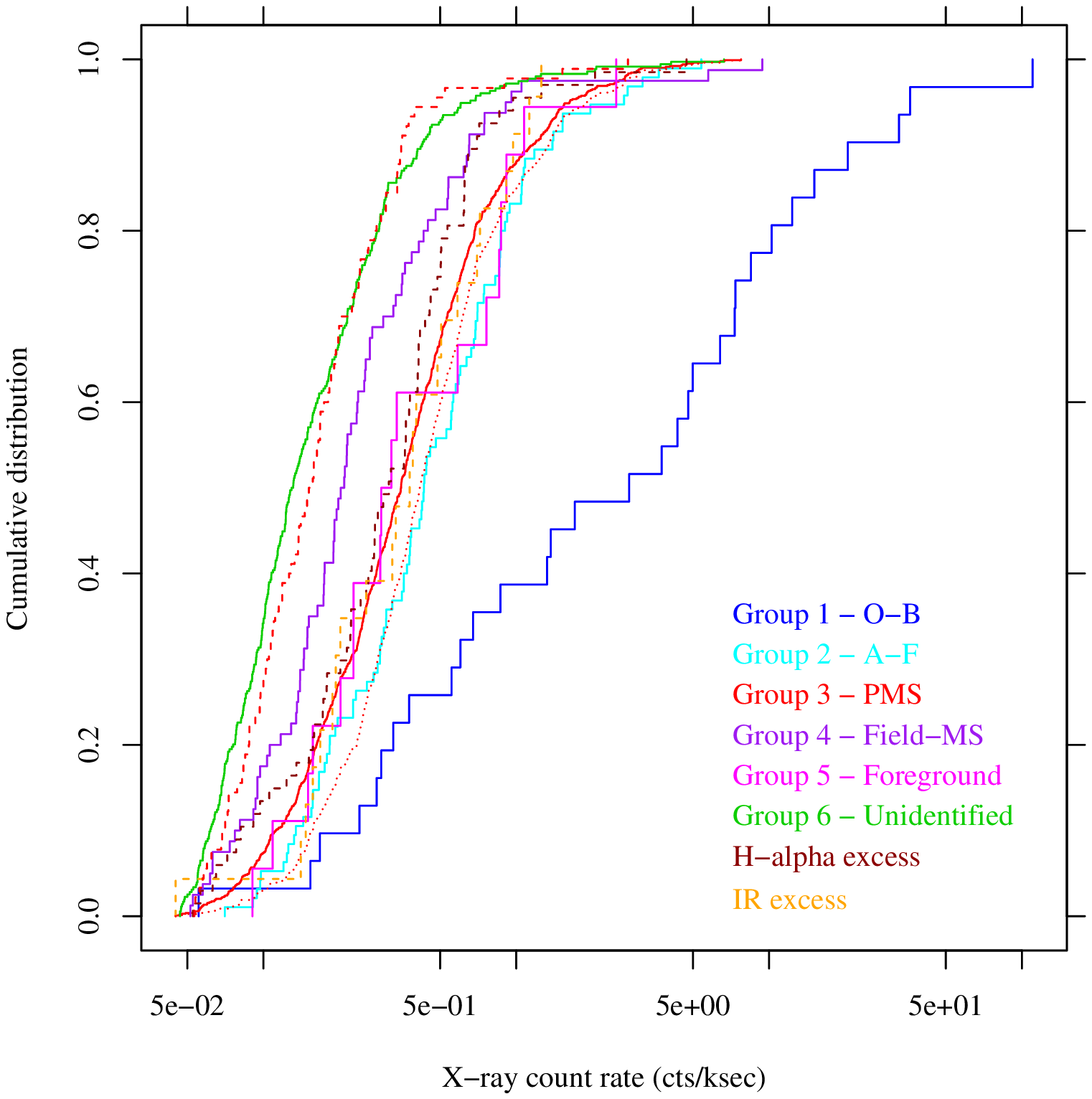}}
\caption{Cumulative distributions of time-averaged X-ray count rates for
the various groups. The PMS group is shown in its entirety (solid red
line), and split between stars respectively brighter (dotted red line)
and fainter (dashed red line) than $V=20.5$.
\label{cumfun-xrate}}
\end{figure}

We study here how X-ray emission is related to stellar properties in
NGC~6231. Apart from massive stars, very little detail is known
for other cluster members. For these latter
we use the available optical photometry to
estimate stellar masses and bolometric luminosities $L_{bol}$, assuming
for all stars the same average cluster reddening and distance.
In particular, masses and $L_{bol}$ were computed for PMS stars using the BHAC
or Siess \e (2000)
5-Myr isochrone in the $(V,V-I)$) CMD (as discussed in Sect.~\ref{cmd}),
and masses of higher-mass MS stars (above $\sim 7 M_{\odot}$) using
the 3-Myr non-rotating solar-metallicity
isochrone from Ekstr\"om \e (2012).
The B0.7Ia supergiant HD~152235 was assigned a mass of 30~$M_{\odot}$
according to Fraser \e (2010).
Bolometric luminosities for OB stars were taken from Sana et al.\ (2006b).
For X-ray sources showing flares in their lightcurve, we have
interpolated or extrapolated the lightcurve outside the flare to compute
``quiescent'' X-ray luminosities, which are used here (both quiescent
and average $L_X$ values are reported in Table~\ref{table1}, as well as the
Kolmogorov-Smirnov test probability $P_{K-S}$ that the source emission
is not compatible with being constant).

Figure~\ref{lx-mass-ir} shows the X-ray luminosity $L_X$ vs.\ the
stellar mass for cluster member stars.
At low masses, $L_X$ increases rapidly with mass until $M \sim
2 M_{\odot}$, shows a large spread between $2 M_{\odot} < M < 10
M_{\odot}$, and increases again for higher masses. We have seen in
Section~\ref{cmd} that the $2-3 M_{\odot}$ range is the one comprising
stars either just arrived on the ZAMS, or still crossing the radiative
track during their PMS evolution. This relatively small mass range
therefore comprises a wide range of stellar spectral types (from type A
to G-early~K),
because of the cluster age spread, and a corresponding range
of stellar structures (from 
absence to preponderance of convective envelopes). The large $L_X$ spread 
we find for the $2-3 M_{\odot}$ range agrees with a different origin between
X-rays from early and late-type stars in this mass range,
with late-type stars being
more X-ray luminous, up to $L_X \sim 10^{31}$ erg/s. The mass range in
which the PMS-ZAMS transition occurs depends on the cluster age (and age
spread): for a younger cluster such as the ONC this transition occurs at higher
masses, so that the X-ray luminosity keeps increasing with mass up to
higher masses ($\sim 3 M_{\odot}$) while still in the PMS stage, and reaches
up to $L_X \sim 10^{31.5}$ erg/s or more (Flaccomio et al.\ 2003),
before it declines in higher-mass ZAMS stars.
As already noted when discussing the \ha-index diagram of
Fig.~\ref{v-rha-xray}, the pattern of $L_X$ vs.\ stellar mass of
Fig.~\ref{lx-mass-ir} agrees therefore with the estimated age of NGC~6231.

Figure~\ref{cumfun-xrate} shows cumulative distributions of X-ray
count rates for the various groups. In order to test if, as suggested
above, the (soft) optically unidentified X-ray sources are consistent
with being low-mass cluster members we have shown the cumulative count-rate
distribution for PMS stars both in its entirety (red solid line), and
separately for stars brighter and fainter than $V=20.5$, respectively
(red dotted and dashed lines): the latter count-rate distribution is nearly
identical to that of unidentified sources, strengthening even more our
previous arguments in favor of their membership.

X-ray emission from A-type stars, which we detect in a large number, is a
puzzle because these stars, most of which already on the ZAMS
or in their late PMS stages
(Fig.~\ref{v-vi-xray}), do not possess deep convection zones. When
detected, X-ray emission from A stars is often attributed to lower-mass
binary companions, an hypothesis which we can readily test.
In Fig.~\ref{profile-ir-xray-2} we have seen that detected A-F stars
have a similar spatial distribution to detected PMS stars; moreover,
their distributions of HR1 and HR2 were also similar to those of the PMS stars
(Figs.~\ref{hr1-distr-ir-groups} and~\ref{hr2-distr-ir-groups}),
consistently with the binary hypothesis. Here, Fig.~\ref{cumfun-xrate}
shows that their count-rate distribution is almost identical to that of the
(brighter) PMS stars, again consistent with the binary hypothesis.
Finally, the number ratio between the X-ray detected A-F stars and the
undetected A-F stars (which were argued above to be members because of
their spatial distribution) is 95/189 = 0.5, which corresponds to a
binary fraction of 0.33 if binarity accounts for all X-ray detections in
the A-F group. This number is not unreasonable, especially in NGC~6231
which was found to contain a large percentage of massive binaries; it is
also consistent with our (lower limit to the) binary fraction for PMS
stars found above.
Sana \e (2006b) also conclude that the binary picture is
the best explanation for their X-ray detection of late-B and A stars in NGC~6231.

\begin{figure}
\resizebox{\hsize}{!}{
\includegraphics[bb=5 10 485 475]{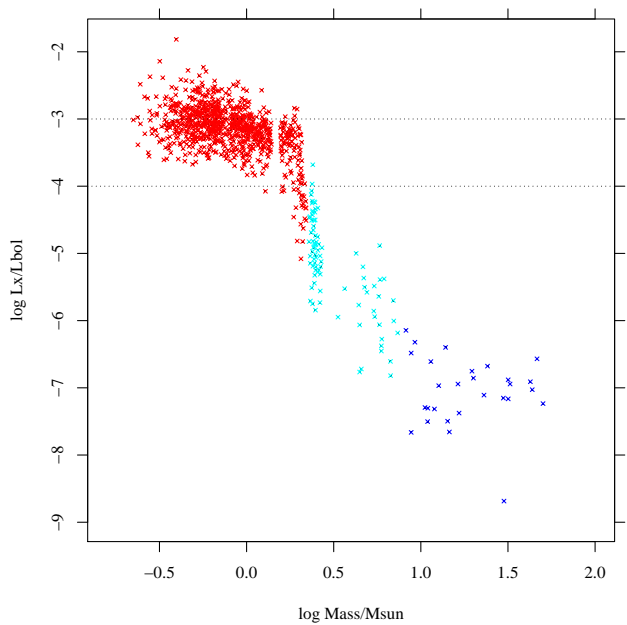}}
\caption{Ratio $L_X/L_{bol}$ vs.\ stellar mass.
The two dotted lines indicate $\log L_X/L_{bol} =-3$ and $-4$, respectively.
\label{lxlbol-mass-ir}}
\end{figure}

We also show in Figure~\ref{lxlbol-mass-ir} the dependence of the ratio
$L_X/L_{bol}$ vs.\ stellar mass, already studied for e.g.\ the ONC
(Flaccomio et al.\ 2003, their Fig.6) and NGC~2362 (Damiani et al.\
2006a, their Fig.13).
The bulk of NGC~6231 PMS stars fall in the range $-3.5 < \log L_X/L_{bol}
< -2.5$. This is a higher average $L_X/L_{bol}$ ratio than PMS stars
in the ONC in the same mass range (Flaccomio et al.\ 2003).
Since it has been shown that ONC PMS stars
with strong accretion from a circumstellar disk have on average a
reduced X-ray emission with respect to non-accreting stars of the same
mass, the observed behaviour in NGC~6231 is consistent with the low fraction
of cluster stars having accretion disks ($\sim 5$\%, see Section~\ref{nir}),
and stars with strong \ha-emission ($\sim 9$\%, see Section~\ref{halpha}).
For more massive stars ($2 < \log L_{bol}/L_{bol,\odot}
< 6$) our result agrees qualitatively with that of Sana et al.\ (2006b).
In both the ONC and NGC2362 a nearly constant $L_X/L_{bol}$ ratio is found
for low masses up to a cutoff mass, beyond which the ratio decreases rather
abruptly by several orders of magnitude, down to the canonical ratio
$\sim 10^{-7}$ for the OB stars. Fig.~\ref{lxlbol-mass-ir}
shows that this age-dependent discontinuity occurs at a mass $\sim 2
M_{\odot}$ in NGC~6231, while the corresponding Figure for the ONC
(1-2~Myr old) shows a cutoff mass of $3 M_{\odot}$. In NGC~2362 (5~Myr)
the cutoff mass is similar to that of NGC~6231, but is less well defined
since it is a less populated cluster.

\begin{figure}
\resizebox{\hsize}{!}{
\includegraphics[bb=5 10 485 475]{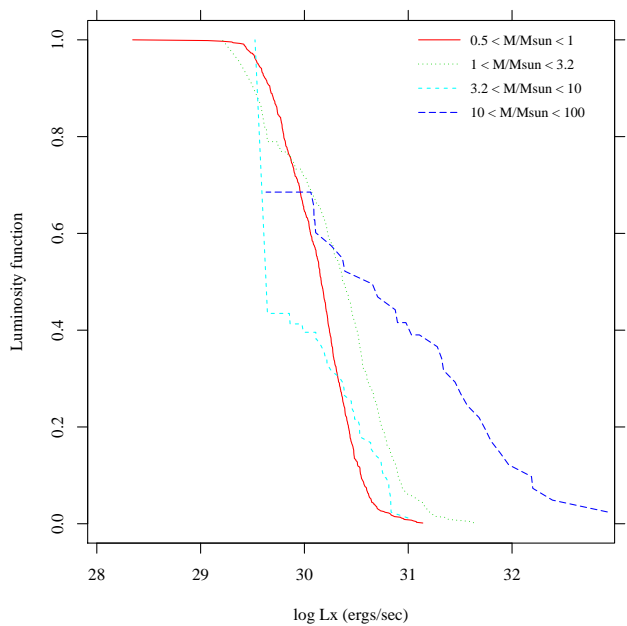}}
\caption{Kaplan-Meier maximum-likelihood X-ray luminosity functions for
members stars in different mass ranges, as indicated.
\label{xlumfun-ir-mass}}
\end{figure}

Figures~\ref{lx-mass-ir}, \ref{cumfun-xrate} and \ref{lxlbol-mass-ir}
are based on X-ray selected stellar samples, and thus
do not account for X-ray undetected member stars.
We have therefore computed X-ray luminosity upper limits for the
undetected cluster members in groups OB and A-F, to build
maximum-likelihood $L_X$ distributions, separately for different mass
ranges, as shown in Figure~\ref{xlumfun-ir-mass}.
We have included among members the optically unidentified soft X-ray sources,
which according to the discussion above are members near the low mass
end of the X-ray detected PMS members.
Since our X-ray selected sample of PMS
members is complete down to a limiting flux
$L_X \sim 8 \times 10^{29}$ erg/s, we infer
from Figure~\ref{lx-mass-ir} that this corresponds to almost 100\%
completeness in mass down to $\sim 1 M_{\odot}$.
We expect a number of X-ray non-detections in the lowest mass range shown
in Fig.~\ref{xlumfun-ir-mass}; however we did not include $L_X$ upper
limits in that range, because we do not have an unbiased member list there.
The red distribution in Fig.~\ref{xlumfun-ir-mass} is therefore an
overestimate of the true distribution in the corresponding mass range.
From Fig.~\ref{xlumfun-ir-mass} we see that in the $3-10 M_{\odot}$ mass
range we do not
sample the median X-ray luminosity, by a small margin, because of the
large fraction of non-detections.

\section{The cluster initial mass function}
\label{imf}

\begin{figure}
\resizebox{\hsize}{!}{
\includegraphics[bb=5 10 485 475]{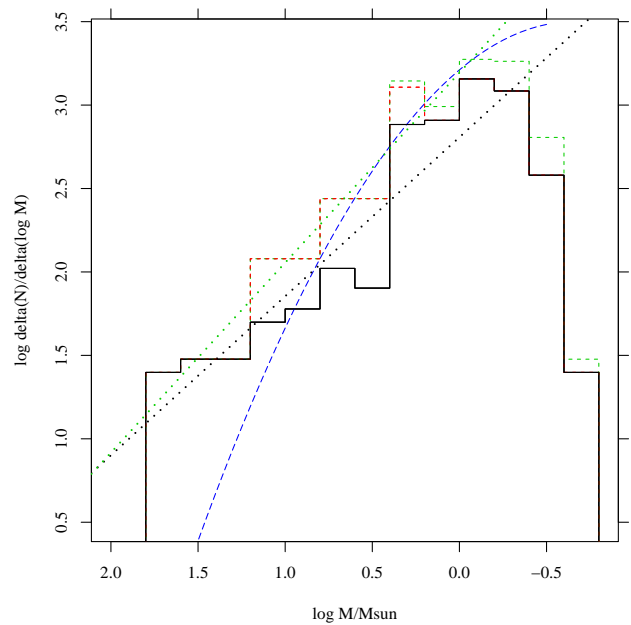}}
\caption{Initial mass function (IMF) of NGC~6231 stars, computed under
different assumptions on star membership, as explained in the text.
Dotted lines are linear best fits to the black IMF (black dotted line)
and to the green one (green dotted line).
The blue dashed line is the Chabrier (2003) model IMF, with arbitrary
normalization.
\label{imf-ir-himass}}
\end{figure}

\begin{figure}
\resizebox{\hsize}{!}{
\includegraphics[bb=5 10 485 475]{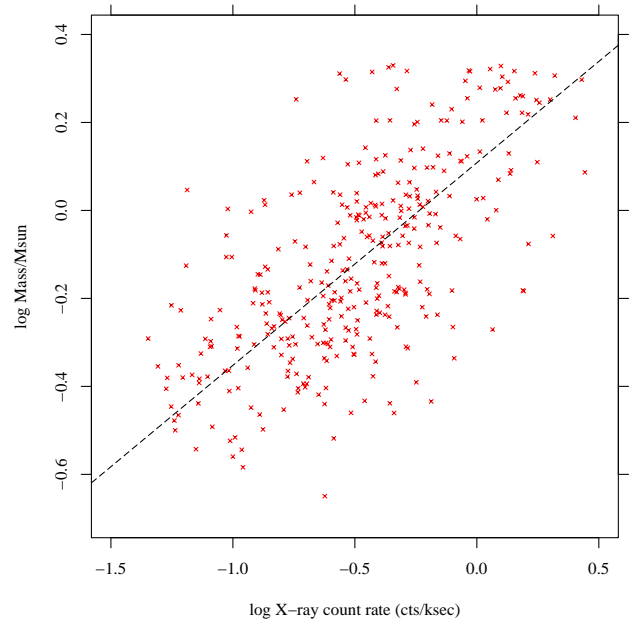}}
\caption{Mass vs.\ X-ray count rate for PMS members.
The dashed line is obtained from a principal-component analysis.
Only non-flaring X-ray sources are shown, within 4~arcmin from cluster center.
\label{mass-xrate}}
\end{figure}

On the basis of the NGC~6231 membership determined above, we have also
computed the cluster initial mass function (IMF).
In order to have a better understanding of the uncertainties involved in
the IMF determination, we have done it under several different
assumptions.
The first, and most conservative, approach is to consider only X-ray
detected stars in groups OB, A-F, and PMS to be reliable members; as we
have argued above, this is expected to yield very incomplete members samples
among the B- and A-type stars, so the resulting IMF is a sort of ``lower
limit'' to the true IMF. The IMF obtained in this way is shown in
Figure~\ref{imf-ir-himass}, as the black solid line. A linear fit in the
plotted quantities (black dotted line)
gives $\log [ \Delta N/\Delta \log M ] \sim 2.81  -0.95 \log M$
(with star mass $M$ in solar-mass units), including only mass bins above
$\log M = 0$, which is our estimated completeness limit as derived
above.

A better assumption is to include among members also X-ray undetected stars
in groups OB and A-F, at the risk of including a (minor) fraction of field
stars as contaminants. It might be argued that the exact fraction of
field stars in these color ranges can be estimated from consideration of
nearby control fields; this approach is not assumption-free, however,
since we argued above of the existence of strongly-rising extinction
just behind the cluster, which might not be present in a ``clean''
reference field, i.e.\ sufficiently far from NGC~6231 center not to contain
cluster members. We choose therefore not to rely on a reference field,
and accept all OB and A-F candidate members in building an ``upper
limit'' IMF: this is shown by the red dashed line in
Fig.~\ref{imf-ir-himass}, sensibly different from the first IMF for
masses between 3-10 $M_{\odot}$.

Finally, we presented above strong evidence that the soft, optically
unidentified X-ray sources must be low-mass members as well; lacking
optical data their mass must be estimated differently. Here we take
advantage of the $L_X$-mass empirical relation found e.g.\ in
Fig.~\ref{lx-mass-ir} (as in many other young clusters). Expressed in
terms of the X-ray count rate (which makes the derivation independent of
the adopted conversion factor between count rate and flux), we replot it
in Figure~\ref{mass-xrate}, for PMS stars only.
A principal-component analysis yields the relation
$\log M/M_{\odot} \sim 0.1084 + 0.46 \; \log r_X$,
where $r_X$ is the ACIS X-ray count rate in counts/ksec.
Adding these mass estimates for unidentified soft X-ray sources,
as well as the few \ha-emission X-ray
undetected members, to the previous IMF yields the one shown in
Fig.~\ref{imf-ir-himass} with a green dashed line: the increment is
modest, and almost limited to the lowest five mass bins, most of which are
anyway affected by incompleteness, as discussed above.
A linear fit gives $\log [ \Delta N/\Delta \log M ] \sim 3.19 (\pm 0.12)
-1.14 (\pm 0.12) \log M$
(green dotted line).
The blue long-dashed line is instead the Chabrier (2003) model, with arbitrary
normalization, which is not a good approximation of our data.

We therefore adopt $\Gamma = -1.14$ as the ``best'' value for the IMF
slope, as this is based on the most complete NGC~6231 member list.
SSB obtain from their analysis of NGC~6231 photometry a value of $\Gamma
=-1.1$, slightly less steep than ours
but consistent within the error,
for $\log M/M_{\odot} \geq
0$. We suspect that this is affected, at the low-mass end, by an
excessive statistical subtraction of putative field stars, based on a
control field. Supporting this suggestion is the
different sensitivity to faint stars near the cluster core and in any
neighboring field free of bright stars, as already discussed in previous
sections.
Our X-ray based membership is very robust for the PMS members, down to
our completeness limit which is nearly coincident with the lower-mass limit
of the SSB IMF. Comparison between the CMD source density in the PMS
band and that in the adjacent Field-MS and Foreground zones suggests that
only very few (perhaps 1\%) of the X-ray PMS members may be
contaminants.

The total mass in our complete IMF is $2.81 \times 10^3 M_{\odot}$
(of which $2.28 \times 10^3 M_{\odot}$ above $1 M_{\odot}$),
slightly larger than the SSB estimate ($2.6 (\pm 0.6) \times 10^3 M_{\odot}$),
but consistent with it within the uncertainties.
Since the true IMF is unlikely to drop abruptly below our detection
threshold, the true total mass must be larger than the above value.
The Weidner \e (2010) analytical IMF formulation
predicts that the total mass is 1.92 times larger than the mass above
$1 M_{\odot}$.
Assuming this ratio holds here as well, and on the basis of our measured
mass above $1 M_{\odot}$,
we may infer a total NGC~6231 stellar mass of $4.38 \times 10^3 M_{\odot}$.
This is very close to the total NGC~6231 mass reported in
Weidner \e (2010) ($4595^{-2312}_{+4676} M_{\odot}$), on the basis of
earlier observations.

\subsection{Completeness and contamination}
\label{complete}

\begin{figure}
\resizebox{\hsize}{!}{
\includegraphics[bb=5 10 485 475]{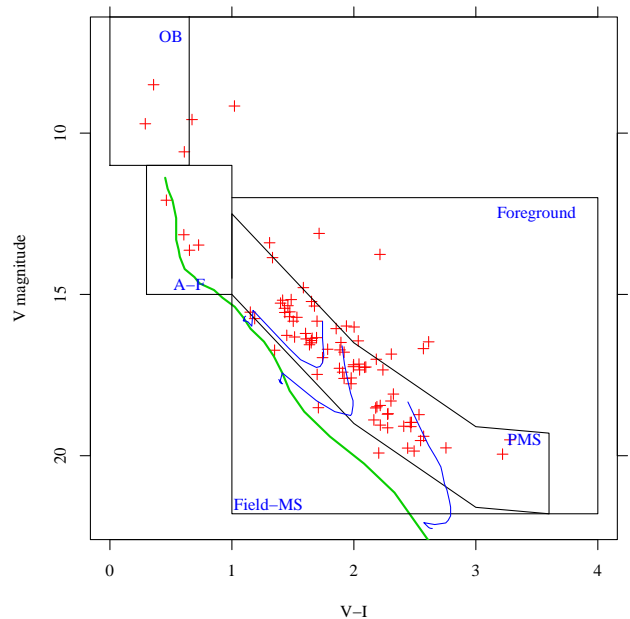}}
\caption{
Optical CMD of XMM-Newton X-ray sources falling outside our Chandra FOV
(red crosses). The same loci as in Fig.~\ref{v-vi-xray} are indicated in
black. The green line is the ZAMS at the cluster distance, while in blue
are BHAC evolutionary tracks for stars of masses $M=0.5, 1, 1.4 M_{\odot}$.
\label{v-vi-xmm}}
\end{figure}

We mention here two competing effects which may have biased our above
determinations of the IMF shape and total cluster mass: the number of
cluster members falling outside the studied field (incompleteness), and the 
number of unrelated young stars mistakenly included in our sample
(contamination). This latter might be important because NGC~6231 belongs to
the Sco~OB1 association (see the review by Reipurth 2008). The space
density of stars in NGC~6231 is clearly much higher than the diffuse
population of Sco~OB1, and we argued above that most of the cluster is
comprised in our Chandra ACIS FOV, so these effects are small corrections
to our determinations. We may nevertheless obtain an estimate of the number
of stars involved from the existing XMM-Newton observations of Sana \e
(2006a, 2007). There are 132 of their X-ray sources falling outside our combined
ACIS FOV, over a spatial region $\sim 1.4$ times wider than our surveyed area.
Their placement in the optical CMD is shown in Figure~\ref{v-vi-xmm}.
Two ot the most massive stars there are included by Sana \e (2008) among
NGC~6231 members. Most of the lower-mass additional X-ray sources
in the Figure fall in the same PMS band as our studied members, and are
therefore good candidates as additional NGC~6231 members (110 stars),
distributed
in a low-density halo around the cluster core. This is consistent with the
spatial density at the outer border of our FOV ($\leq 1$ source/square arcmin,
Fig.~\ref{clust-profile3}). The total combined mass of all presumably
missed members is of order of $\sim 150 M_{\odot}$, or an additional 7\% to
our above estimates. A fraction of them
might however belong to the diffuse Sco~OB1 population, which is little
studied and therefore difficult to quantify, especially with respect to the
low-mass populations. If we for example assume that the low-mass
stars detected with XMM-Newton are genuine NGC~6231 halo members, but the few OB
stars are from the diffuse population (and the same number of extraneous OB
stars must be subtracted from our sample), we would add $\sim 100
M_{\odot}$ of low-mass stars and subtract $\sim 50 M_{\odot}$ of OB stars,
with a net increase of only $\sim 50 M_{\odot}$ in our total-mass estimate
($\sim 2$\%). Therefore, those estimates are not severely affected by
incompleteness and contamination. It may be interesting to remark that if
the last scenario is true (a few contaminating OB stars and many missed
halo low-mass members in our sample), this might at least partially
account for the low IMF slope found above.

\section{Conclusions}
\label{concl}

We have observed with Chandra/ACIS-I the young cluster NGC~6231 in
Sco~OB1, detecting 1613 point X-ray sources down to $\log L_X \sim
29.3$ (a complete sample down to $\log L_X \sim 29.8$). Most of the
detected sources are without doubt low-mass cluster members, and permit
an unprecedented study of the late-type population of this very rich
cluster. The cluster morphology is found to be nearly spherical, with a
small elongation orthogonal to the galactic plane. Comparison of the
X-ray source list with existing optical and 2MASS catalogues has allowed
identification of 85\% of the X-ray sources.
The optical color-magnitude diagram of X-ray detected sources shows,
in addition to many tens massive stars,
a richly populated PMS cluster band
with a definite spread in ages (between $\sim$1-8~Myr).
The PMS stars appear slightly
older than the massive (main-sequence or evolved) OB stars.
Upon comparison with existing model isochrones, we find the Baraffe \e
(2015) models in better agreement with our data than the older Siess \e
(2000) models. The shape of the $V$-magnitude spread across the PMS
band is suggestive of a (equal-mass) binary fraction of $\sim 20$\%.
The IR color distribution suggests a rapid increase in extinction, by
several magnitudes in $V$, at some distance behind the cluster.

The likelihood of membership for different star groups, defined from
their position on the optical CMD, was studied by means of their spatial
distribution, X-ray hardness and luminosity. We argue that the vast
majority of OB and A-F stars are cluster members, regardless of their
X-ray detection. The X-ray detection fraction of late-B and A stars is
consistent with the hypothesis that the actual X-ray emitter is a
later-type companion, while the more massive star is X-ray dark.
We confirm the spatial
segregation of the most massive stars, as suggested by previous studies.

We find 203 \ha-excess stars in the X-ray FOV, with \ha\ intensities
typical of CTTS. About one-half of them lie in the cluster PMS band, and
are added to our member list; of the latter, $\sim 40$\% were not
detected in X-rays. The derived fraction of \ha-excess PMS stars is
$\sim 9$\%, in good agreement with the trend shown by other clusters at
the same age. The \ha-excess stars have no significantly different
X-ray properties than the rest of PMS stars.
We also find 48 bona-fide IR-excess sources, probably related to circumstellar
disks, whose frequency is therefore $\sim 5$\% for NGC~6231 PMS stars.
The spatial distribution of the IR-excess stars, both X-ray detected and
undetected, is significantly wider than that of PMS stars: we argue that
this is evidence of increased photoevaporation of disks in the vicinity
of the cluster OB stars. A similar trend is also shown by X-ray
undetected CTTS, strengthening this conclusion.

Approximately 70\% of the optically unidentified X-ray sources show
(spatial and X-ray) properties well consistent with those of the cluster
PMS stars at their low-mass end, and we suggest that they are low-mass
members as well. The remaining 30\% of unidentified sources show instead
much harder, multi-component X-ray spectra very unlike cluster stars,
and do not appear otherwise connected with the cluster.
With more than 1300 stars with individually-determined membership, this
dataset enables us to study NGC~6231 with an unprecedented detail, across
a wide range of masses, down to $\sim 0.3 M_{\odot}$ (complete to $\sim
1 M_{\odot}$).

The X-ray properties of NGC~6231 stars are in agreement with those of
coeval clusters. The mass above which $L_X/L_{bol}$ drops is an age
indicator, which agrees with the isochronal age, and is also consistent with
the age-dependent discontinuity in the $(H\alpha,V)$ diagram.

We find a complex, doubly-peaked X-ray spectrum for the
Wolf-Rayet star WR~79, with more extreme characteristics than
exhibited by stars of similar type; WR~79 is one of the few WC-type stars
detected in X-rays.

Using all available membership information, we compute the cluster IMF
under different hypotheses, and over nearly 2 orders of magnitude in
mass. We find a best-fit slope
$\Gamma = -1.14$, consistent with
that found by SSB but slightly closer to the Salpeter value.
The total mass of
individually selected cluster members above $1 M_{\odot}$ is found to be 
$2.28 \times 10^3 M_{\odot}$, while the total mass extrapolated to the
bottom of the mass spectrum according to the Weidner \e (2010)
analytical IMF is $4.38 \times 10^3 M_{\odot}$,
in agreement with
previous estimates. This confirms that NGC~6231 ranks among the
most massive known young clusters in the Milky Way.

\begin{appendix}

\section{Near-IR extinction map}
\label{extinct-map}

\begin{figure*}
\centering
\includegraphics[width=18cm]{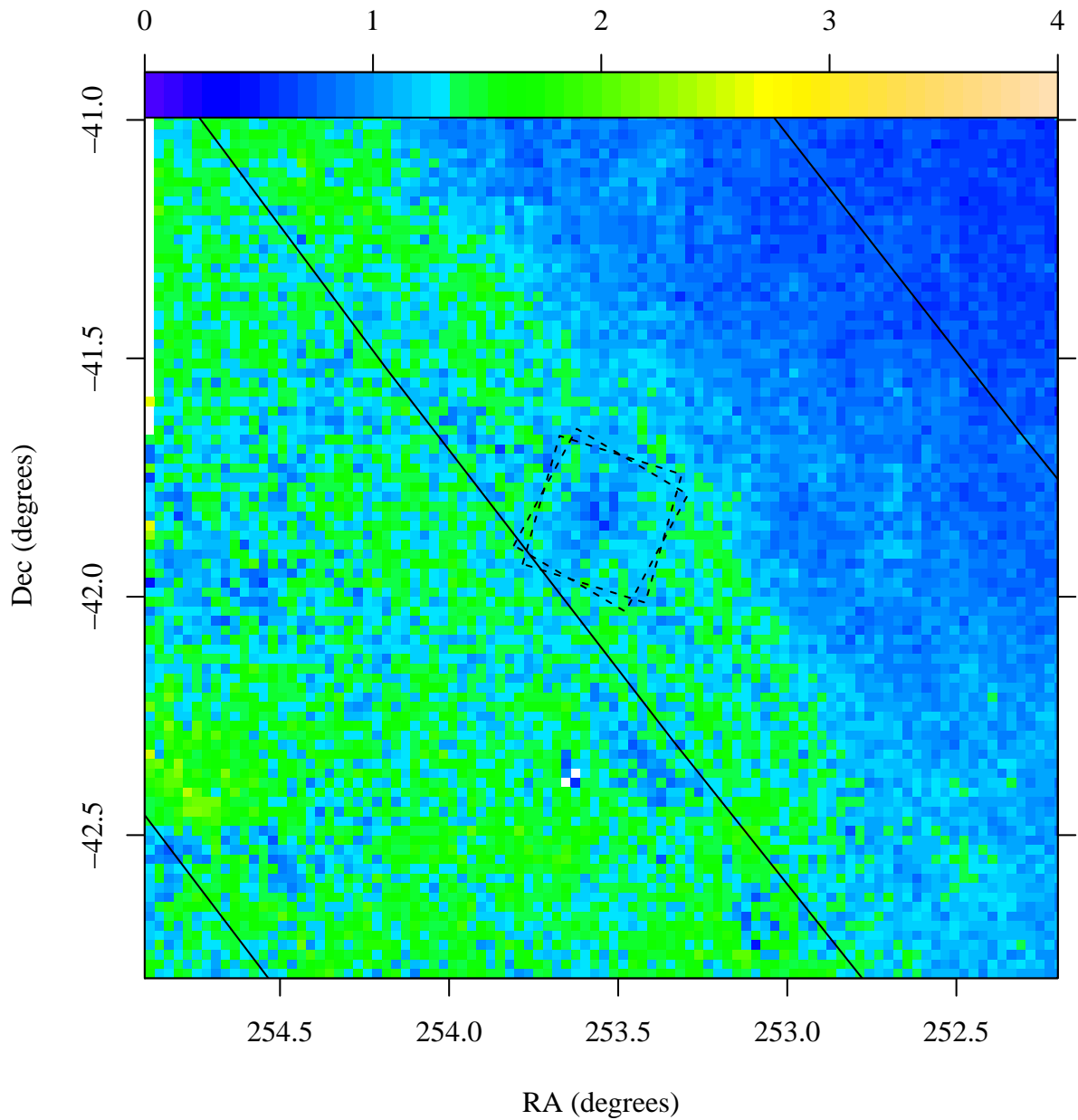}
\caption{
Spatial map of average $J-H$ color, over a $2^{\circ} \times
2^{\circ}$ region around NGC~6231. The top-axis colorbar and labels indicate
the correspondence between displayed color and $J-H$.
The two Chandra FOVs are indicated with dashed squares. The
three solid lines indicate, left to right, galactic latitudes $b=0,1,2$
degrees, respectively.
\label{spatial-jh-mean}}
\end{figure*}

\begin{figure*}
\centering
\includegraphics[width=18cm]{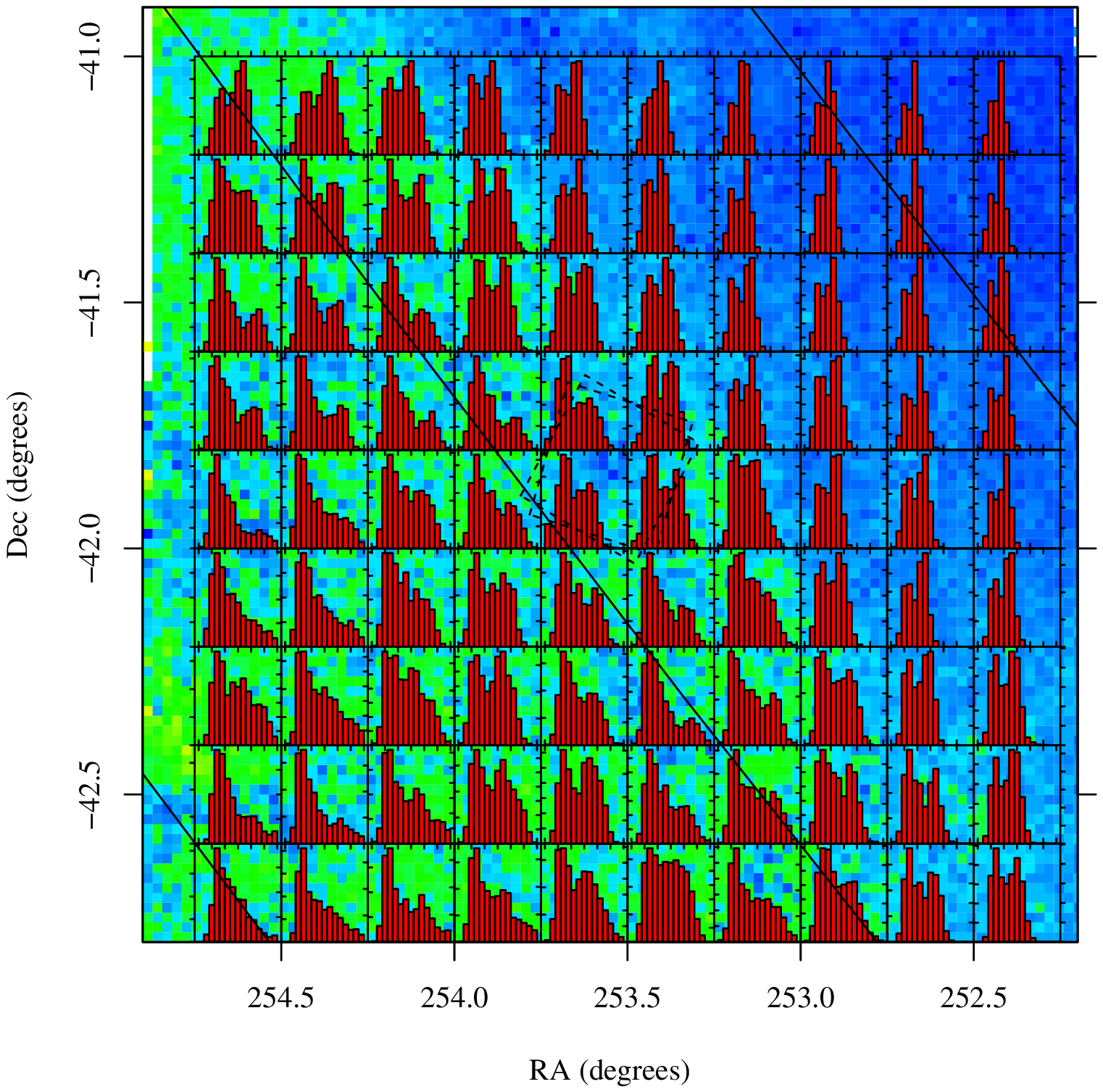}
\caption{
Same field and background image as in Fig.~\ref{spatial-jh-mean}, with the
addition of local $J-H$ histograms. Each histogram covers a $J-H$ range of
$[0,3]$.
\label{spatial-jh-hist}}
\end{figure*}

In order to understand better the environment in which NGC~6231 is located,
we examine here the distribution of $J-H$ colors of stars over a wider area
($2^{\circ} \times 2^{\circ}$). Since the intrinsic stellar $J-H$ falls in
the small range $[0,3]$, this permits to estimate roughly the extinction
towards the stars within the 2MASS sample in the area. The corresponding
spatial map is shown in Figure~\ref{spatial-jh-mean}, based on more than
400,000 2MASS point sources. At large distances, dominant contributors are
red giants, whose intrinsic $J-H$ colors are always close to $\sim 1$
(Bessell and Brett 1988). Assuming a ratio $E(J-H)/A_V=0.107$ (Rieke and
Lebofsky 1985) also permits to infer optical extinction values from this
map. It is seen that NGC~6231 is surrounded by a region of higher average
extinction, and that this increases systematically towards the galactic
plane, to the south-east.

However, this is only an oversimplified description, since the true
three-dimensional distribution of absorbing material has considerable
structure. This can be seen from examining the local distribution of $J-H$
(on regions $10^{\prime} \times 12^{\prime}$ in size): these local
histograms are shown in Figure~\ref{spatial-jh-hist}, over the same spatial
region as above. While at $b \sim 2$ the distribution has usually a single
peak, at the latitudes near NGC~6231 ($b \sim 1$), including the cluster
region itself, it is very often bimodal. Near $b \sim 0$ an extended red
tail is often seen in place of the red peak.
The bimodality in the color distribution is an indication of a rapid
increase in extinction over a short spatial distance; when the extinction
increases smoothly no second peak is produced, like in the lowest-$b$
histograms. In the neighborhood of the cluster, the two $J-H$ peaks are
separated by $\Delta (J-H) \sim 0.7$ corresponding to $\Delta A_V \sim 7$.
The dust density enhancements producing this extra extinction must be located
behind the cluster, since this is rather uniformly extincted, as seen above.
We cannot determine exactly how far this absorbing layer is from the cluster,
but it might be
related to residual dust from its parental molecular cloud, and therefore
not very far behind it.

Some additional constraint may be obtained from the highly-reddened,
background yellow supergiant V870~Sco (F5Ia, Herbig 1972).
This star has apparent $V=12.4$ and $B-V=3.4$ (SSB, star 3926);
based on its spectral type and
luminosity class, the intrinsic color is $(B-V)_0=0.26$ (FitzGerald 1970),
and the color excess $E(B-V)=3.14$. Assuming $R_V \sim 3.2$, visual extinction
is $A_V=10.0$, and the extinction-corrected $V_0=2.4$. Again based on
spectral type, the absolute visual magnitude is $M_V = -8.2 \pm 1$
(Sung \e 2013b),
and the distance modulus $(m-M)=10.6 \pm 1$ ($1320^{+770}_{-490}$~pc).
Herbig (1972) remarked that not only
this star is much more reddened than NGC~6231 members, but has also a
radial velocity incompatible with membership, and is likely a background
object, which sets its minimum distance modulus to 11.0.
The extinction difference between V870~Sco and NGC~6231 ($\Delta
A_V \sim 8.5$) must therefore originate between the cluster distance (1585~pc)
and the distance of this star (up to 2090~pc, $(m-M)=11.6$), i.e.\
along 500~pc or less.
This is a much more rapid $A_V$ rise than between us and NGC~6231, in
agreement with our considerations above. We are in debt with the referee
for calling our attention to this star.

\end{appendix}

\begin{acknowledgements}
We thank the referee for his/her useful suggestions.
We acknowledge support from the italian MIUR.
The scientific results reported in this article are based
on observations made by the Chandra X-ray Observatory (ObsIds: 5372,
6291).
This research has made use of software provided by the Chandra X-ray
Center (CXC) in the application package CIAO.
This work was also using data products from observations made with
ESO Telescopes at the
La Silla Paranal Observatory under programme ID 177.D-3023, as part of
the VST Photometric \ha\ Survey of the Southern Galactic Plane and Bulge
(VPHAS$+$, www.vphas.eu).
This study has made use of the 2MASS database.
We also used data from the SIMBAD database, operated at CDS, Strasbourg.
\end{acknowledgements}

\bibliographystyle{aa}

\end{document}